\documentclass[aps,twocolumn,amsmath,amssymb,reprint,numbers,superscriptaddress,noeprint]{revtex4-1}

\usepackage{graphicx}
\usepackage{dcolumn}
\usepackage{bm}
\usepackage{bbm}
\usepackage{soul}
\usepackage{enumitem}
\usepackage{makecell}

\DeclareMathOperator{\sgn}{sgn}

\usepackage{braket}
\usepackage{graphicx}
\usepackage{amsmath,verbatim,latexsym,amssymb,indentfirst,mathrsfs,mathtools,amsthm,bbm,bm,hyperref,url}

\usepackage[title,titletoc]{appendix}

\newcommand{\Bell}{{\rm B}}  

\newcommand*{\ketbra}[2]{\lvert #1 \rangle\!\langle #2 \rvert}

\newcommand*{\expval}[1]{\left\langle  #1  \right\rangle}
\usepackage{color}
\usepackage{bbold}
\usepackage[dvipsnames]{xcolor}
\usepackage[normalem]{ulem}
\usepackage{physics}
\newcommand{\kater}[1]{{\color{black}#1}}

\newcommand{\nicole}[1]{{\color{black} #1}}

\begin{document}

\title{Agnostic Phase Estimation}

\author{Xingrui Song}
\affiliation{
Department of Physics, Washington University, St. Louis, Missouri 63130, USA
}

\author{Flavio Salvati}
\affiliation{
Cavendish Laboratory, Department of Physics, University of Cambridge, Cambridge, CB3 0HE, United Kingdom
}

\author{Chandrashekhar Gaikwad}
\affiliation{
Department of Physics, Washington University, St. Louis, Missouri 63130, USA
}

\author{Nicole Yunger Halpern}
\affiliation{
Joint Center for Quantum Information and Computer Science, NIST and University of Maryland, College Park, Maryland 20742, USA
}
\affiliation{
Institute for Physical Science and Technology, University of Maryland, College Park, Maryland 20742, USA
}

\author{David R. M. Arvidsson-Shukur}
\affiliation{
Hitachi Cambridge Laboratory, J. J. Thomson Avenue, Cambridge CB3 0HE, United Kingdom
}

\author{Kater Murch}
\affiliation{
Department of Physics, Washington University, St. Louis, Missouri 63130, USA
}

\date{\today}

\begin{abstract}
The goal of quantum metrology is to improve measurements' sensitivities by harnessing quantum resources.  Metrologists often aim to maximize the quantum Fisher information, which bounds the measurement setup's sensitivity. In studies of fundamental limits on metrology, a paradigmatic setup features a qubit (spin-half system) subject to an unknown rotation. One obtains the maximal quantum Fisher information about the rotation if the spin begins in a state that maximizes the variance of the rotation-inducing operator. If the rotation axis is unknown, however, no optimal single-qubit sensor can be prepared.  Inspired by simulations of closed timelike curves, we circumvent this limitation. We obtain the maximum quantum Fisher information about a rotation angle, regardless of the unknown rotation axis. To achieve this result, we initially entangle the probe qubit with an ancilla qubit. Then, we measure the pair in an entangled basis, obtaining more information about the rotation angle than any single-qubit sensor can achieve. We demonstrate this metrological advantage using a two-qubit superconducting quantum processor. Our measurement approach achieves a quantum advantage, outperforming every entanglement-free strategy.
\end{abstract}

\maketitle

Phase estimation is crucial to
quantum information 
\nicole{processing: In several algorithms, phase \kater{estimation} identifies unitary operators' eigenvalues~\cite{Shor94,Brassard02,Bauer20,Suzuki20,Lloyd20,Lloyd21}. }
Furthermore, phase estimation 
\nicole{serves} in quantum metrology, the 
\nicole{use of} quantum systems to 
\nicole{probe and estimate physical parameters} \cite{Braunstein94, Giovanetti06, Giovanetti11}. 
Conventionally, phase estimation requires 
\nicole{prior} knowledge about the unitary being 
\nicole{probed.}

For example, consider a unitary $e^{i {H}}$ generated by a Hamiltonian ${H}$. In quantum algorithms, phase estimation encodes  in a qubit register an estimate of an $e^{i {H}}$ eigenvalue \cite{Kitaev95,Nielsen11}. To perform this encoding, one 
\nicole{initializes} another 
\nicole{register} in an $H$ eigenstate. Without information about ${H}$, conventional algorithmic phase estimation fails. 

In quantum metrology, phase estimation is 
\nicole{used} to infer some unknown parameter $\alpha$
in a unitary ${U}_\alpha=e^{i \alpha {A}}$. 
The Hermitian generator ${A}=\sum_i a_i \ketbra{a_i}{a_i}$ has eigenstates $\ket{a_i}$ and eigenvalues $a_i$.   
\nicole{$\alpha$} could quantify an unknown field's strength. 
One can estimate $\alpha$ by applying ${U}_\alpha$ to several quantum systems and measuring them. The 
optimal single-qubit probe states are the equal-weight superpositions of the 
\nicole{$\ket{a_i}$} associated with the greatest and least eigenvalues, e.g., $ \left( \ket{a_{\mathrm{min}}} + \ket{a_{\mathrm{max}}}  \right) / \sqrt{2}$ \cite{Braunstein94, Giovanetti06, Giovanetti11}. The 
optimal measurement observables depend on $A$, too. Without information about ${A}$, therefore, conventional  metrological phase estimation  fails.

If ${H}$ or ${A}$ is unknown, one can \nicole{first}
learn about it through quantum-process tomography~\cite{Chuang97, DAriano01, Altepeter03, Song2021}.
However, process tomography requires many applications of $e^{i {H}}$ or ${U}_\alpha$, 
plus many measurements. The number of applications of a unitary quantifies resource usage in quantum computing and metrology. 
Hence tomography is costly. Furthermore,
one often cannot 
\nicole{leverage} process tomography. 
For example, consider a magnetic field whose direction changes. 
We might wish to measure the field strength $\alpha$ at some instant. The probes must be prepared optimally beforehand.

\nicole{Recently,~\cite{ArvidssonShukur23}} outlined a phase-estimation protocol for when information about ${A}$ becomes available \textit{after} the unitary \nicole{operates.} 
The protocol harnesses the mathematical equivalence between certain entanglement-manipulation experiments and closed timelike curves, hypothetical worldlines that travel backward in time \cite{Godel49, Morris88, Bennett05, Svetlichny11, Lloyd11, Lloyd11-2}. In the protocol of~\cite{ArvidssonShukur23}, one entangles a probe 
\nicole{and} ancilla. After information about ${A}$ becomes available, one effectively updates the probe's initial state, by measuring the ancilla, using the equivalence.  
\nicole{This prescription inspires} metrological protocols that leverage entanglement to circumvent requirements of 
\nicole{prior} information. Optics experiments have explored the relationship between entanglement manipulation and closed timelike curves \cite{Ringbauer2014, Marletto2019}. Additionally, delayed-choice quantum-erasure experiments \cite{Kaiser2012,Lee2014} resemble the protocol in~\cite{ArvidssonShukur23} conceptually. However, metrological protocols inspired by closed timelike curves have not been \nicole{reported.} 

\nicole{We} show that entanglement manipulation can enable optimal estimation of $\alpha$, even sans information about $A$.   
We consider a common scenario: an arbitrary unbiased estimator $\hat{\alpha}$ is calculated from $N$ measurement outcomes.  
The \emph{Fisher information} (FI) $I_\alpha$ quantifies the outcome probabilities' sensitivity to small changes in $\alpha$. 
\nicole{$I_\alpha$, defined below,} limits the estimator's variance through the \emph{Cram\'er--Rao bound}:  
${\rm var}({\hat \alpha}) \geq 1/(N I_\alpha)$.
We theoretically prove that entanglement can boost the FI of
$\alpha$ by $50 \, \%$. 
Our protocol is optimal, achieving the FI of the optimal protocol that leverages knowledge of $A$. Using 
\nicole{superconducting qubits,} we demonstrate the advantage experimentally. 


The next sections present and experimentally demonstrate four strategies for inferring about $\alpha$ without knowledge of the rotation axis $\hat{\bm{n}}$. A \emph{single-qubit sensor} can extract no information about $\alpha$. Two time-travel-inspired protocols follow: \emph{hindsight sensing} consumes a maximally entangled two-qubit state. The protocol achieves an FI of 1 if information about $\hat{\bm{n}}$ becomes available eventually.  \emph{Agnostic sensing}  requires a maximally entangled two-qubit state and an entangling measurement. The protocol achieves an FI of 1 even if $\hat{\bm{n}}$ remains unknown. We compare these entanglement-boosted protocols to \emph{entanglement-free sensing with an ancilla} 
\nicole{whose FI is $2/3$.}

\begin{figure}
    \centering
    \includegraphics[width=8.6cm]{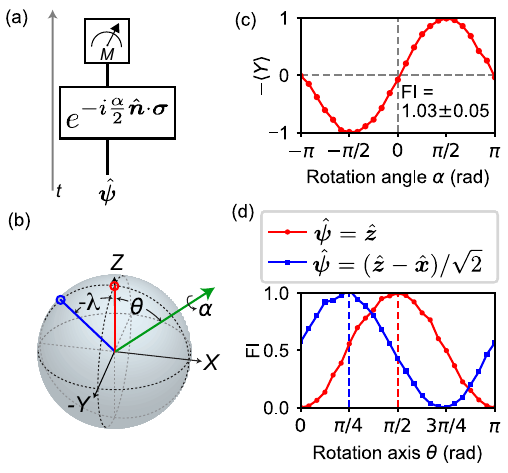}
    \caption{{\bf Fisher information achievable 
    \nicole{with} single-qubit sensor.} (a) Protocol for sensing the rotation angle $\alpha$. Time runs \nicole{vertically,}
    as in the closed time-like-curve representation introduced later. 
        (b)  Bloch-sphere representation of the protocol. The red and blue lines represent possible initial states. The green arrow indicates the rotation axis. 
        (c) Outcomes from preparing 
        $\hat{\bm{\psi}} = \hat{\bm{z}}$, then rotating about the $\hat{\bm{x}}$-axis through a varying rotation angle $\alpha$. The red points and curve represent the measured $-\expval{Y}$ values, from which we infer the FI ($\pm$ one standard deviation).
        (d) FI measured at various rotational axes parameterized by $\theta$. If the initial state is $\hat{\bm{\psi}} = \hat{\bm{z}}$ (red curve), the FI fails to achieve its maximum value, except if $\theta = \pi/2$ specifies the rotation axis. An analogous statement concerns $\hat{\bm{\psi}} = (\hat{\bm{z}}-\hat{\bm{x}})/\sqrt{2}$ 
        (blue curve) and $\theta = \pi/4$.}
    \label{fig:single_qubit}
\end{figure}

\emph{Single-qubit quantum sensor.}---\nicole{The simplest quantum sensor is}
a qubit probe subject to an 
\nicole{unknown} rotation, represented \nicole{by} 
$U_{\bm{\alpha}} = \exp \left(-i\alpha\hat{\bm{n}}\cdot\bm{\sigma} / 2 \right)$.
The unknown rotation angle is $\alpha$, 
$\hat{\bm{n}} = \sin \theta \cos \phi \, \hat{\bm{x}} 
+ \sin \theta \sin \phi \, \hat{\bm{y}} 
+ \cos \theta \, \hat{\bm{z}}$ defines the 
unknown rotation axis, and $\bm{\sigma} = (X, Y, Z)$ denotes a vector of Pauli operators. \kater{$\theta$ and $\phi$ denote the polar and azimuthal \nicole{angles.} } 
Figure~\ref{fig:single_qubit}(a) illustrates the protocol: the probe is prepared in $\ket{\psi}$, 
\nicole{evolves} to $\ket{\psi_{\alpha}} \coloneqq U_{\alpha} \ket{\psi}$, 
and is 
\nicole{measured projectively.}
\nicole{One aims to} infer the rotation angle $\alpha$.  

Consider measuring the probe in an arbitrary basis $\{ \ket{i} \}$. Outcome $i$ occurs with probability $P_i = |\langle i | \psi_{\alpha}\rangle|^2$. The FI quantifies these probabilities' $\alpha$-sensitivity:
$I_\alpha = \sum_{i = 0,1}{\frac{(\partial_{\alpha} P_i)^2}{P_i}} $. The FI is upper-bounded by the \emph{quantum Fisher information} (QFI), $\mathcal{I}_\alpha$~\cite{Braunstein94,helstrom76}:
\begin{equation}
        I_\alpha \leq
    \mathcal{I}_\alpha = 4 \, \left( \langle \partial_\alpha \psi_{\alpha} | \partial_\alpha \psi_{\alpha}\rangle - | \langle \psi_{\alpha} | \partial_\alpha \psi_{\alpha}\rangle |^2 \right)  . 
    \label{eq_QFI_bound_form}
\end{equation}

The QFI, 
\nicole{itself,} is upper-bounded by the  maximum variance of 
\nicole{the generator $\hat{\bm{n}}\cdot \hat{\bm{\sigma}} /2$ of $U_\alpha$:}
$\mathcal{I}_\alpha 
\leq 4 \max_{\ket{\psi}} \{ \mathrm{var}(\hat{\bm{n}}\cdot \hat{\bm{\sigma}} /2) \} = 1$.  
Consider the 
\nicole{many-trial limit} (as $N \rightarrow \infty$). If $\hat{\bm{n}}$ is known, all bounds (including the 
\nicole{introduction's Cram\'er--Rao bound)}
can be saturated: 
\begin{equation}
\label{Eq:Bounds}
    {\rm var}({\hat \alpha}) 
    = \frac{1}{ N I_{\alpha} }
    =  \frac{1}{ N \mathcal{I}_{\alpha} }
    =  \frac{1}{4 N \max_{\ket{\psi}} \left\{ \mathrm{var} \left( \frac{ \hat{\bm{n}}\cdot \hat{\bm{\sigma}} }{2} \right) \right\}} = \frac{1}{N} .
\end{equation}
If $\hat{\bm{n}}$ is unknown, neither saturation happens, typically~\cite{supp}.

Figure~\ref{fig:single_qubit}(b) depicts the protocol on the Bloch sphere. 
Without loss of generality, 
\nicole{the rotation axis lies} in the $\hat{\bm{x}}$--$\hat{\bm{z}}$-plane. 
We choose the pure initial state's Bloch vector to be $\hat{\bm{\psi}} = \sin(\lambda)\hat{\bm{x}} \ + \cos (\lambda) \hat{\bm{z}}$, illustrating 
\nicole{diverse} metrological outcomes, from worst to optimal. 
Figure~\ref{fig:single_qubit}(b) shows two possible initial states, with $\lambda =0 $ (red) and $\lambda = -\pi/4$ (blue). 
The Supplemental Material~\cite{supp} details 
\nicole{single-qubit rotations' implementation.}
The rotation 
\nicole{evolves the probe state to} $\ket{\psi_\alpha} $. 
Our later analysis governs all
$\alpha \in [-\pi, \pi]$,
but illustrating with infinitesimal 
\nicole{rotations} $\dd  \alpha$ first is instructive.
An infinitesimal rotation displaces the Bloch vector by an amount
$\dd \bm{\psi} = \hat{\bm{n}} \times \hat{\bm{\psi}}\ \dd \alpha = \sin (\lambda - \theta) \hat{\bm{y}}\ \dd\alpha 
= \hat{\bm{y}}\ \dd\expval{Y}$. 
Therefore, an optimal final measurement is of $Y$.

Figure~\ref{fig:single_qubit}(c) displays data from optimal measurements.
We show $-\expval{Y}$ at multiple $\alpha$ values for the initial state $\lambda = 0$ and rotation-axis parameter $\theta = \pi/2$.  We 
\nicole{calibrated} and corrected for the $\approx 98\ \%$ measurement fidelity throughout this work~\cite{supp}. 
Fitting the outcomes to a sinusoid, we infer the $P_i$s at $\alpha=0$. From 
\nicole{them,} we calculate $I_\alpha = 1.03\pm0.05$, 
\nicole{consistently} with the maximum predicted QFI.

Figure~\ref{fig:single_qubit}(d) displays the measured FI for various rotation axes. The red curve (initial state parameterized by $\lambda=0$) shows that the maximum FI is achieved only when $\theta=\pi/2$.  The blue curve (initial state 
\nicole{with} $\lambda=-\pi/4$) shows a maximum FI only at $\theta = \pi/4$. 
These results \kater{illustrate} 
\nicole{an above-mentioned point:} one can generally obtain the maximum 
\nicole{FI} about 
\nicole{$\alpha$} only 
if 
\nicole{prior} knowledge about $\theta$ informs the probe's preparation and measurement.
This limitation betrays a deeper problem with the single-qubit \nicole{probe~\cite{supp}: }
$U_\alpha$ has three unknown parameters, whereas a qubit has two degrees of freedom (DOFs).
Estimating $\alpha$, 
\nicole{without knowing the rotation axis,} is therefore typically impossible.

\begin{figure}
    \centering
    \includegraphics[width=8.6cm]{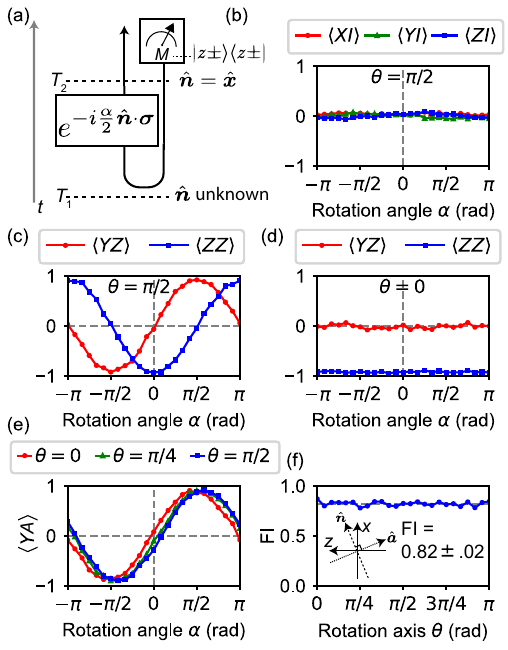}
    \caption{
        {\bf 
        \nicole{Hindsight sensing.} }
        (a) Protocol for sensing the rotation angle $\alpha$ by mimicking a closed timelike curve. The ancilla's state effectively travels backward in time. It flips at $T_1$, becoming an optimal probe state.
        (b) 
        \nicole{No probe observable's expectation value} carries information about $\alpha$. 
        (c) If the rotation is about the $x$-axis (if $\theta = \pi / 2$), the probe--ancilla correlators $\expval{YZ}$ and $\expval{ZZ}$ encode information about $\alpha$. 
        (d) 
        \nicole{For} different rotations about the $z$-axis ($\theta = 0$), the same correlators contain no information about $\alpha$. 
        (e) The correlator $\expval{YA}$ depends on the optimal ancilla observable to measure.  
        $\expval{YA}$ is sensitive to the rotational angle $\alpha$, at rotation axes parameterized by $\theta=0, \pi / 4$, and $\pi / 2$. 
        (f) From the correlator, we calculate the FI, for various rotational axes. The FI remains close to the optimal value, 1. The subfigure indicates the optimal ancilla measurement axis $\hat{\bm{a}}$.
    }
    \label{fig:two_qubits}
\end{figure}

\emph{Hindsight sensing.}---\nicole{We}
relax the requirement of 
\nicole{prior} knowledge about 
\nicole{ $\hat{\bm{n}}$, harnessing } 
the connection between closed timelike curves and entanglement~\cite{Lloyd11,lloyd_11,Brun2017,allen_14}.  
Consider preparing a maximally entangled (Bell) state 
\nicole{[the} $\cup$ symbol in Fig.~\ref{fig:two_qubits}(a)] between \emph{two} qubits at 
\nicole{time} $T_1$. One can view this preparation as the chronology-violating trajectory of \emph{one} qubit that travels backward in time, turns around at $T_1$, and continues forward in time \cite{Bennett05, Svetlichny11, Lloyd11, Lloyd11-2}. 
We harness this connection to effectively choose a probe's initial state \emph{in hindsight}. 

Figure~\ref{fig:two_qubits}(a) illustrates this strategy. At $T_1$, we initialize a probe qubit and an ancilla qubit in a singlet. A unitary $U_{\alpha}$ rotates the probe's state  about an unknown axis $\hat{\bm{n}}$. Afterward, $\hat{\bm{n}}$ is revealed; Eq.~\eqref{Eq:Bounds} can be satisfied. 
We measure the ancilla along an axis orthogonal to $\hat{\bm{n}}$.
The measurement projects the ancilla onto an optimal rotation-sensing state. 
The probe's state 
\nicole{becomes} orthogonal to the ancilla's. 
One can imagine that the time-traveling qubit in Fig.~\ref{fig:two_qubits}(a) is flipped at $T_1$. Hence \nicole{closed timelike curves inspire our experiment.}

In previous metrology protocols \cite{ArvShukur20,ArvShukur21, Lupu22,Salvati_23}, 
\nicole{an ancilla measurement} determined whether the probe would undergo a final, information-acquiring measurement. Our protocol always features 
\nicole{probe and ancilla measurements.} The ancilla-measurement outcomes 
help us postprocess the data from  probe-measurement outcomes  to infer about $\alpha$.

\nicole{All four Bell states~\cite{Nielsen11} } serve 
\nicole{equally well,}
we prove in~\cite{supp}. We illustrate with the \nicole{singlet,} 
whose effectiveness 
\nicole{we} understand intuitively through the state's rotational invariance:
\begin{equation}
    \ket{\Psi^{-}} = \dfrac{1}{\sqrt{2}}\left( \ket{b}_\mathrm{P} \ket{\bar{b}}_\mathrm{A} - \ket{\bar{b}}_\mathrm{P} \ket{b}_\mathrm{A} \right) .
\end{equation}
$\mathrm{P}$ denotes the probe; and $\mathrm{A}$, the ancilla. The structure of $\ket{\Psi^{-}}$ does not depend on the single-qubit basis $\{ \ket{b}, \ket{\bar{b}} \}$; $\ket{\Psi^{-}}$ remains invariant under identical rotations of $\mathrm{P}$ and $\mathrm{A}$.
\nicole{Denote} by $\ket{a_0}$ and $\ket{a_1}$ the 
$-\hat{\bm{n}}\cdot \hat{\bm{\sigma}} /2$ eigenstates associated with the eigenvalues $+\frac{1}{2}$ and $-\frac{1}{2}$. Define the superpositions 
$\ket{a^{\pm}} \equiv (\ket{a_0} \pm \ket{a_1} )/\sqrt{2}$. 
\nicole{Measuring the ancilla's $\{ \ket{a^\pm} \}$ projects the probe }
onto an optimal state for measuring $\alpha$. 

Figure~\ref{fig:two_qubits} details this protocol's experimental implementation. Using a parametric entangling gate, we prepare the probe and ancilla in a singlet~\cite{supp}. We then rotate the probe and perform tomography on the probe--ancilla state.
Figure~\ref{fig:two_qubits}(b) displays the 
\nicole{measured probe expectation values}
when $\theta=\pi/2$ parameterizes the rotation axis. These expectation values encode no information about $\alpha$, the flat curves indicate. This lack is expected, since 
\nicole{each qubit's reduced state is} maximally mixed. 

To learn about $\alpha$, we 
\nicole{calculate} two-qubit correlators.
Figure~\ref{fig:two_qubits}(c) illustrates with $\langle YZ \rangle$. 
\nicole{Using entanglement, we reproduce}
the results of Fig.~\ref{fig:single_qubit}(c): measuring the ancilla's $Z$ projects the probe's Bloch vector onto $\pm\hat{\bm{z}}$, which are 
\nicole{optimal} for sensing $\alpha$. However, the sensor's sensitivity depends on the rotation axis. $\expval{YZ}$ and $\expval{ZZ}$ cannot register rotations about the $\hat{\bm{z}}$-axis ($\theta = 0$), Fig.~\ref{fig:two_qubits}(d) shows.

We interpret these results using 
\nicole{closed-timelike-curve language} \footnote{
\nicole{We} do not precisely simulate a (postselected) closed timelike curve~\cite{Lloyd11}: we 
\nicole{discard no} information through postselection. However, such curves motivate and provide intuition about our experiment~\cite{ArvidssonShukur23}. 
}.
When the qubits are initialized in a singlet at $T_1$, the probe is configured agnostically: 
for every axis $\hat{\bm{m}}$, $\langle \bm{\sigma} \cdot \hat{\bm{m}} \rangle = 0$.
The probe is waiting for the optimal-state input from the future. The probe 
\nicole{is rotated;} and the ancilla's optimal basis, $ \{ \ket{a^\pm} \}$, is measured at $T_2$. 
The measurement projects the ancilla's state onto $\ket{a^{\pm}}$.
This state is effectively sent backward in time and flipped into $\ket{a^{\mp}}$, to serve as the probe's time-$T_1$ state.
Thus, the probe is retroactively prepared in the optimal state; is rotated with $U_{\alpha}$; and, at $T_2$, undergoes a $Y$ measurement.

Figure~\ref{fig:two_qubits}(e) demonstrates that we can obtain the maximum FI by measuring the ancilla 
\nicole{$\hat{\bm{n}}$-dependently,} as by measuring $\{ \ket{a^\pm} \}$.
Figure~\ref{fig:two_qubits}(f) displays the FI obtained when $\theta \in [0,\pi]$ parameterizes the rotation axis. Regardless of the axis, we obtain a QFI of $\approx 0.82$.  This value is less the maximum possible QFI, due to 
\nicole{the entangled-state preparation's finite fidelity~\cite{supp}. } 

\begin{figure}
    \centering
    \includegraphics[width=8.6cm]{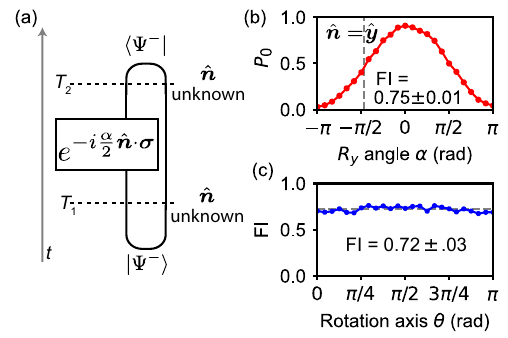}
    \caption{
        {\bf Agnostic sensing.}  
         (a) Protocol: The probe and ancilla are prepared in a singlet. The probe is rotated, whereupon we measure whether the qubits 
         \nicole{remain} in the singlet.
        (b) $P_0$ denotes the probability of obtaining 
        \nicole{a \emph{yes}. } From $P_0$, we infer the FI. 
        Different plots follow from rotations about the $\hat{\bm{x}}$-, $\hat{\bm{y}}$-, and $\hat{\bm{z}}$-axes. 
        (c) FI inferred after various rotations in the $\hat{\bm{x}}$--$\hat{\bm{z}}$ plane. }
    \label{fig:singlet_singlet}
\end{figure}

\emph{Agnostic sensing}.---The previous section's protocol 
\nicole{lets us} effectively choose the probe's initial state after $U_\alpha$.
\nicole{Yet} an entangling measurement, beyond the entangled initial state, enables an optimal sensing strategy that requires neither 
\nicole{prior} nor 
\nicole{later} knowledge of the rotation axis: what we term an \emph{agnostic sensor}. Similar protocols have been studied in different \nicole{settings~\cite{lloyd_11,giro14,Yuan2016}. }

Figure~\ref{fig:singlet_singlet}(a) sketches the protocol. 
\nicole{Again,} we initialize the probe and ancilla in a singlet, $\ket{\Psi^{-}}$. 
\nicole{The} probe undergoes an unknown rotation $U_{\alpha}$,
\nicole{which} maps $\ket{\Psi^{-}}$ to 
\begin{align}
    \ket{\Psi^{-}_\alpha}  = \frac{1}{\sqrt{2}} \Big( e^{+i\frac{\alpha}{2}} \ket{a_0}_\mathrm{P} \ket{a_1}_\mathrm{A} + e^{-i\frac{\alpha}{2}} \ket{a_1}_\mathrm{P} \ket{a_0}_\mathrm{A} \Big).
\end{align}
Finally, we perform an entangling measurement of
$\{ \Pi_0 \coloneqq \ketbra{\Psi^-}{\Psi^-}, \,
\Pi_1 \coloneqq \mathbb{1} - \Pi_0 \}$.
The possible outcomes' probabilities are
\begin{align}
   P_0  & \coloneqq \left\lvert \bra{\Psi^{-}_{\alpha}} \Pi_0 \ket{\Psi^{-}_{\alpha}} \right\rvert^2  
   = \cos^2 (\alpha/2) \; \: \text{and}
   \\ \quad 
    P_1 & \coloneqq \left\lvert \bra{\Psi^{-}_{\alpha}} \Pi_1 \ket{\Psi^{-}_{\alpha}}  \right\rvert^2 
   = \sin^2 (\alpha/2).
\end{align}
This strategy produces the maximum FI about $\alpha$, regardless of the rotation axis \cite{Braunstein94}. 
More broadly, Eq.~\eqref{Eq:Bounds} holds. 

We can understand this result through closed timelike curves.
If $\alpha=0$, then $U_{\alpha}$ does not perturb the initial state $\ket{\Psi^-}$, and $P_0 =1$. 
The ancilla--probe pair maps onto a particle traversing a closed timelike curve infinitely many times~\cite{Lloyd11,Lloyd11-2}.
If $\alpha \neq 0$, 
\nicole{$U_{\alpha}$ } imprints $\alpha$ on the state.
\nicole{This} experiment has a probability $P_0 < 1$ of successfully simulating a closed timelike curve.
Knowing this success probability enables us to estimate $\alpha$.

Figure~\ref{fig:singlet_singlet} details our experimental 
\nicole{demonstration.}
We prepare the singlet via an $\sqrt{i\mathrm{SWAP}}$ gate~\cite{supp}.
The probe then undergoes $U_{\alpha}$.
\nicole{We} measure $\{ \Pi_0, \Pi_1 \}$  by rotating  the ancilla, performing another $\sqrt{i\mathrm{SWAP}}$, (this process maps the singlet onto separable states), and measuring 
\nicole{each qubit's $Z$ eigenbasis.}
From many 
\nicole{trials,} we infer $P_0$.

We measure $P_0$ for rotations about the $\hat{\bm{x}}$-, $\hat{\bm{y}}$-, and $\hat{\bm{z}}$-axes, for several $\alpha$ values [Fig.~\ref{fig:singlet_singlet}(b--d)]. As expected, $P_0 \propto \cos^2(\alpha/2)$, independently of the rotation axis.  We fit $P_0$ near $\alpha=-\pi/2$  to infer the FI. Figure~\ref{fig:singlet_singlet}(e) displays the measured FI for rotation axes $\theta \in [0,\pi]$: regardless of the axis, $\mathcal{I}_{\alpha} = 0.72$. 

The two-qubit gate's fidelity limits the \nicole{FI,} 
as in 
\nicole{hindsight sensing. Agnostic sensing }
requires two such gates, so the infidelity impacts the FI more. This proportionality highlights 
\nicole{a} trade-off between quantum advantage and circuit depth.

\emph{Entanglement-free sensing with ancilla.---}\nicole{Let us compare the agnostic sensor }
with an optimal entanglement-free sensor.
Imagine restricting the probes to identical single-qubit pure states. All pure states serve equally \nicole{well,}
by symmetry: the rotation axis is unknown.
Without loss of generality, therefore, we suppose 
\nicole{the probes} begin in $\ket{\psi} = \ket{0}$. 
Three independent, unknown parameters specify $U_\alpha$:
the rotation angle $\alpha$, 
\nicole{plus} the rotation axis's zenith angle $\theta$ and azimuthal angle $\phi$.
One cannot estimate 3 parameters using a qubit, whose state encodes only 2 DOFs. For every single-qubit probe, 
\nicole{some $(\alpha,\theta,\phi)$ values yield}
$\mathcal{I}_{\alpha} = 0$.  Hence, no single-qubit probes achieve Eq.~\eqref{Eq:Bounds}, we prove in~\cite{supp}.

Nevertheless, one can estimate $\alpha $ without consuming entanglement, e.g., by performing quantum-process tomography on  $U_{\alpha}$ \cite{Ariano2001, Altepeter2003}. 
Since $\hat{\bm{n}}$ is unknown, the most reasonable prior distribution for $\hat{\bm{n}}$ is uniform.
We describe a strategy for garnering the greatest average FI inferable from any entanglement-free input:  
prepare the 
\nicole{probe} in a state $\ket{\psi_j}$, tagged with an ancilla state $\ket{j}$, with probability $p_j$:
\begin{equation}
\rho_0 = \sum_j {p_j \ketbra{\psi_j}{\psi_j} \otimes \ketbra{j}{j}} 
\, , \; \text{wherein} \; 
\sum_j p_j = 1 .
\end{equation}
We show the following in~\cite{supp}. First, for all $\rho_0$, the QFI about $\alpha$, averaged over the $\hat{\bm{n}}$, equals $2/3$. Second, not all $\rho_0$ achieve  the first two equalities in Eq.~\eqref{Eq:Bounds}. 
Third, we derive the form of the states $\rho_0$ for which (i) $\mathcal{I}_{\alpha} = 2/3$ independently of $\hat{\bm{n}}$ and (ii) the first two equalities in Eq.~\eqref{Eq:Bounds} hold. Examples include $\rho_{\star} = (\ketbra{x+}{x+} \otimes \ketbra{1}{1} + \ketbra{y+}{y+} \otimes \ketbra{2}{2} + \ketbra{z+}{z+} \otimes \ketbra{3}{3} )/3 $, where $\ket{x+}$ denotes the eigenvalue-1 $X$ eigenstate, etc. 
Preparing and optimally measuring $\rho_{\star}$ yields a FI of $2/3$ about $\alpha$ 
[Eq.~(C33)], irrespectively of $\hat{\bm{n}}$. \kater{In \cite{supp} we describe our experimental implementation of this entanglement-free strategy, where we achieve an average axis-independent FI of 0.62,  consistent with the theoretical maximum of $2/3$.}

\emph{Discussion.}---In 
\nicole{\cite{Degen_2017},}  the authors distinguish 
a hierarchy of quantum sensors: 
\nicole{some leverage energy-level quantization (type I), 
others leverage quantum coherence (type II), and 
others leverage entanglement (type III).  }
We have introduced a type-III sensor. It achieves an
advantage over the more-classical type-II sensors\kater{, when the resource is the number of times the unitary is applied}. The 50 \% improvement in the QFI weighs against the cost of entanglement manipulation. One cost that we avoid is postselection: we discard no data. All measurement outcomes inform our inference of $\alpha$, despite a known relationship between postselection and closed timelike curves~\cite{lloyd_11,Lloyd11}.

Several 
\nicole{opportunities} suggest themselves. First, our protocol merits \kater{extending to optical~\cite{opticsnote}} and solid-state systems that have concrete metrological applications. Second, our protocol may benefit phase estimation in quantum algorithms. 
Third, our experiment was inspired by the theoretical application of closed timelike curves to metrology~\cite{ArvidssonShukur23}\nicole{---specifically,} 
to weak-value amplification, a technique for sensing interaction strengths~\cite{Dressel_14_Colloquium,Harris_17_Weak,Pang_14_Entanglement,Xu_20_Approaching,Vaidman_88_Aharonov,Duck_89_Sense,Hosten_08_Observation}. One can experimentally implement the application to weak-value amplification or to a more general technique, partially postselected amplification~\cite{Lupu22}.
Fourth, we 
\nicole{anticipate our technique's usefulness}
in metrology subject to time constraints. For example, one may need to measure a time-varying field at 
\nicole{some} instant \cite{Tang18,Wei21,Stankevic23}. 
To date, optimal sensing strategies have required \textit{a priori}  knowledge about the unknown unitary's generator, $A$. Our agnostic protocol entails optimal state preparations and measurements without this knowledge.

\begin{acknowledgements}
\emph{Acknowledgments.}---The authors are thankful for discussions with Aephraim Steinberg and his lab. 
This research was supported in part by grant NSF PHY-2309135 to the Kavli Institute for Theoretical Physics (KITP).
This project was initiated at the KITP program ``New directions in quantum metrology.''
This research was further supported by NSF Grant PHY-2309135 to the Kavli Institute for Theoretical Physics (KITP),  
PHY-1752844 (CAREER), 
NSF QLCI grant OMA-2120757, 
the Air Force Office of Scientific Research (AFOSR)  Multidisciplinary University Research Initiative (MURI) Award on Programmable systems with non-Hermitian quantum dynamics (Grant No.~FA9550-21-1- 0202), ONR Grant No. N00014- 21-1-2630, and by the Gordon and Betty Moore Foundation, grant DOI 10.37807/gbmf11557. The device was fabricated and provided by the Superconducting Qubits at Lincoln Laboratory (SQUILL) Foundry at MIT Lincoln Laboratory, with funding from the Laboratory for Physical Sciences (LPS) Qubit Collaboratory. D.R.M.A.S. acknowledges support from Girton College, Cambridge. F.S. was supported by the Harding Foundation.
\end{acknowledgements}

\newpage



%

\begin{widetext}

\begin{center}
\large
{\bf Supplemental Information for ``Agnostic Phase Estimation''}
\end{center}

\renewcommand{\thesection}{\Alph{section}}
\renewcommand{\thesubsection}{\arabic{subsection}}
\renewcommand{\thesubsubsection}{\roman{subsubsection}}

\makeatletter\@addtoreset{equation}{section}
\def\theequation{\thesection\arabic{equation}}

\section{Experimental demonstration of the optimal entanglement-free sensing strategy}

\kater{Here we test the strategy detailed in the main text for garnering the greatest average FI inferable from any entanglement-free input. Figure~\ref{fig:classical_ancilla} displays our experimental implementation. We prepare:
$$\rho_{\star} = (\ketbra{x+}{x+} \otimes \ketbra{1}{1} + \ketbra{y+}{y+} \otimes \ketbra{2}{2} + \ketbra{z+}{z+} \otimes \ketbra{3}{3} )/3, $$
where $\ket{x+}$ denotes the eigenvalue-1 $X$ eigenstate, etc.  
If the probe is in $\ket{x+}$, the optimal measurement is of $X$; if $\ket{y+}$, then $Y$; and, if $\ket{z+}$, then $Z$. We use the ancilla as a record of the initial probe state, to choose the optimal measurement.}

\nicole{We calculate} the QFI by averaging the QFI values obtained from the probabilistically combined input states: 
$\mathcal{I}_{\alpha} = I_{\alpha} = \frac{1}{3} (\mathcal{I}_{\alpha, \ket{x+}} + \mathcal{I}_{\alpha, \ket{y+}} + \mathcal{I}_{\alpha, \ket{z+}}) = 2/3$. 
We measure the FI for different rotation axes, as when demonstrating the \kater{other} sensing protocols. We achieve an average axis-independent FI of 0.62,  consistently with the theoretical maximum of $2/3$.  

\begin{figure}[h]
    \centering
    \includegraphics[width=8.6cm]{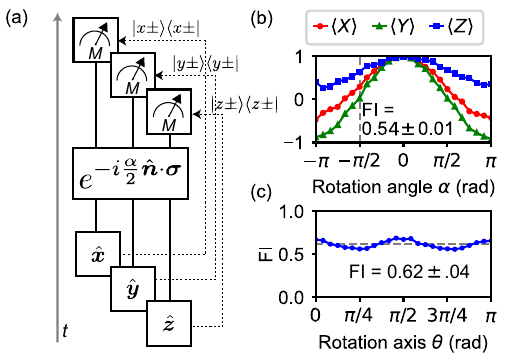}
    \caption{
        {\bf Entanglement-free sensing with ancilla}
         (a) Protocol: Single-qubit-probe states are tagged with classical ancillas that determine the final-measurement basis (dashed lines). 
         (b) For a rotation about the $(\theta = \pi / 5, \phi = \pi / 9)$ axis, different single-qubit-probe states yield different FI values. 
         (c) FI values inferred after various rotations~\cite{sinusoidalnote}. 
         }
    \label{fig:classical_ancilla}
\end{figure}



\section{Experimental setup}

The experimental measurements were made within a two-qubit subsection of a three-qubit device. Further details about the overall setup at the hardware level appear in~\cite{Gaikwad_2024}.  Figure~\ref{fig:setup}(a) displays a simplified schematic of the relevant portion of the device. There are two transmon circuits; one is fixed-frequency, and one is frequency-tunable via a fast flux line (FFL). This tunability allows us to activate parametric entangling gates by modulating the ancilla's frequency [Fig.~\ref{fig:setup}(b)].  In this appendix, we sometimes refer to a transmon circuit's higher-energy eigenstates. We label the energy eigenstates as $\ket{g}$, $\ket{e}$, and $\ket{f}$; and we refer to the circuits as qubits when only the two lowest levels are relevant. The two qubits are coupled via an off-resonant bus resonator. Single-qubit rotations are applied via independent drive lines. The qubits are also coupled to separate readout resonators, which are probed by a common feedline.

\subsection{Dispersive readout}

The probe and ancilla qubits are dispersivley coupled to their respective readout resonators~\cite{Boissonneault2009}, allowing for simultaneous high-fidelity single-shot readout~\cite{Walter2017}. We employ a heterodyne readout scheme: we multiplex the readout signal by simultaneously sending in two pulses that have different frequencies. We separate the two readouts in the frequency domain for processing. For low-noise amplification, we use a traveling-wave parametric amplifier based on the SNAIL (Superconducting Nonlinear Asymmetric Inductive eLements) architecture \cite{Ranadive2022}. The readout fidelity is limited primarily by the narrow cavity bandwidths ($\kappa$, as given in Table \ref{tab:sim}), requiring long integration times. We apply $\pi$-pulses in the $\{\ket{e},\ket{f}\}$-submanifold to increase the  signal-to-noise ratio. 
We optimize over readout amplitude, frequency,  and amplifier-bias
settings with a gradient-free optimization algorithm~\cite{nevergrad}. We determine the
multicomponent-pulse-integration 
envelopes  
via linear-discriminant analysis and principal-component analysis. Ultimately, we achieve readout fidelities of $98.9\ \%$ for the ancilla and $97.8\ \%$ for the probe, utilizing a random forest classifier \cite{scikit-learn}.  After this calibration, all tomography results are corrected for these finite readout fidelities [Fig.~\ref{fig:setup}(c)] via
the iterative Bayesian-update correction method \cite{Nachman2020}. The high-fidelity readout is utilized to implement an active reset protocol for the initial state preparation.

\subsection{Single-qubit rotations}

In this project, we use four types of single-qubit rotations:
\begin{itemize}
    \item Single-qubit $\pi/2$ rotations: The $\pi/2$ rotations applied about the $\hat{\bm{x}}$- and $\hat{\bm{y}}$-axes are used in quantum state tomography and in the arbitrary-axis rotations detailed below. The $\pi/2$ rotations last for
    72 ns (for the ancilla) and 36 ns (for the probe).
    The pulse envelopes are obtained from the convolution of a square wave and two cosine-DRAG-style spectrum filters~\cite{Werninghaus2021}. 
    (DRAG stands for \emph{derivative removal by adiabatic gate}.) We have tuned the two cosine-DRAG-style filters to minimize crosstalk between the two qubits, in addition to minimizing coupling out of the qubit manifolds. The spectral leakage from all the sidelobes is minimized via numerical optimization.  We estimate these gates' fidelity to be 99.6 \%.

    \item Single-qubit $\pi$ rotations: These are implemented in the same manner as the $\pi/2$ rotations. However, the pulse durations are 120 ns (for the ancilla) and 64 ns (for the probe). 

    \item The $\{\ket{e},\ket{f}\}$ 
    rotations are implemented in the same manner as the $\{\ket{g},\ket{e}\}$ rotations, described in the previous two bullet points. 
    The $\pi/2$-gate times are 64 ns (for the ancilla) and 36 ns (for the probe). The $\pi$-gate times are 100 ns (for the ancilla) and 64 ns (for the probe).

    \item Single-qubit arbitrary-axis, arbitrary-angle rotations:
We represent such a rotation with Euler angles: $(\theta_U, \phi_U, \lambda_U)$. The corresponding unitary operator is expressed, relative to the $Z$ eigenbasis, as
\begin{equation}
    U(\theta_U, \phi_U, \lambda_U) =
    \begin{pmatrix}
    \cos \left( \dfrac{\theta_U}{2} \right) & 
    -e^{i\lambda_U} \sin \left( \dfrac{\theta_U}{2} \right)\\
    e^{i\phi_U} \sin \left( \dfrac{\theta_U}{2} \right) & 
    e^{i(\phi_U + \lambda_U)} \cos \left( \dfrac{\theta_U}{2} \right)
    \end{pmatrix}.
\end{equation}
We realize this unitary (up to a global phase) by a combination of rotations,  
\begin{equation}
    \label{Eq_U_as_R_R_RZ}
    U(\theta_U, \phi_U, \lambda_U) = R\left(\frac{\pi}{2}, \phi_U - \pi\right)R\left(\frac{\pi}{2}, \theta_U + \phi_U\right) R_z(\theta_U + \phi_U + \lambda_U),
\end{equation} where $R(\frac{\pi}{2}, \phi)$ is defined as a $\pi/2$ rotation about the $(\cos\phi \, \hat{\bm x} + \sin\phi \, \hat{\bm y})$ axis. A general $R(\alpha, \phi)$, for an arbitrary angle $\alpha$, is expressed as
\begin{equation}
    R(\alpha, \phi) =
    \begin{pmatrix}
    \cos\frac{\alpha}{2} & -i e^{-i \phi}\sin\frac{\alpha}{2} \\
    -i e^{i \phi}\sin\frac{\alpha}{2} & \cos\frac{\alpha}{2}
    \end{pmatrix}.
\end{equation}
We implement $R_z (\alpha)$ (a rotation about the $\hat{\bm{z}}$-axis through an arbitrary angle $\alpha$) through a combination of two $\pi$ rotations:
\begin{equation}
    \label{Eq_RZ_as_R_X}
    R_z(\alpha) = R\left(\pi, \frac{\alpha}{2}\right)R\left(\pi, 0\right).
\end{equation}
The resulting $R_z$ produces a \emph{physical} rotation of the qubit. This fact  
is important, because the agnostic-sensing protocol relies on the accumulation of a physical phase difference between two qubits' states.  Combining Eqs.~(\ref{Eq_U_as_R_R_RZ}) and~(\ref{Eq_RZ_as_R_X}), we implement the arbitrary rotation through
\begin{equation}
    U(\theta_U, \phi_U, \lambda_U) = R\left(\frac{\pi}{2}, \phi_U - \pi\right)R\left(\frac{\pi}{2}, \theta_U + \phi_U\right) R\left(\pi, \frac{1}{2}(\theta_U + \phi_U + \lambda_U)\right)R\left(\pi, 0\right).
\end{equation}
Finally, we relate the Euler angles $(\theta_U, \phi_U, \lambda_U)$ to the arbitrary rotation through an angle $\alpha$, about the axis $\hat{\bm{n}} = \sin \theta \cos \phi \, \hat{\bm{x}} 
+ \sin \theta \sin \phi \, \hat{\bm{y}} 
+ \cos \theta \, \hat{\bm{z}}$.
Given the existence of nonunique solutions, the conversion formulae we define are
\begin{equation}
\begin{aligned}
    \theta_U & = 2\arcsin(\sin\dfrac{\alpha}{2} \sin\theta) , \\
    \phi_U & = \arctan(\tan\dfrac{\alpha}{2} \cos\theta) + \phi - \dfrac{\pi}{2} , \quad \text{and} \\
    \lambda_U & = \arctan(\tan\dfrac{\alpha}{2} \cos\theta) - \phi + \dfrac{\pi}{2} \, .
\end{aligned}
\end{equation}
These formulae are valid for $-\pi \leq \alpha \leq \pi$ and $0 \leq \theta \leq \pi$. 
For the boundary case where $\alpha = \pm \pi$, we define
\begin{equation}
\begin{aligned}
    \theta_U & = \pi - \abs{\pi - 2\theta} \, , \\
    \phi_U & = \sgn(\pi - 2\theta) \dfrac{\pi}{2} + \phi - \dfrac{\pi}{2} \, ,  \quad \text{and} \\
    \lambda_U & = \sgn(\pi - 2\theta) \dfrac{\pi}{2} - \phi + \dfrac{\pi}{2} \, .
\end{aligned}
\end{equation}

\begin{figure}
    \centering
    \includegraphics[width=.9\textwidth]{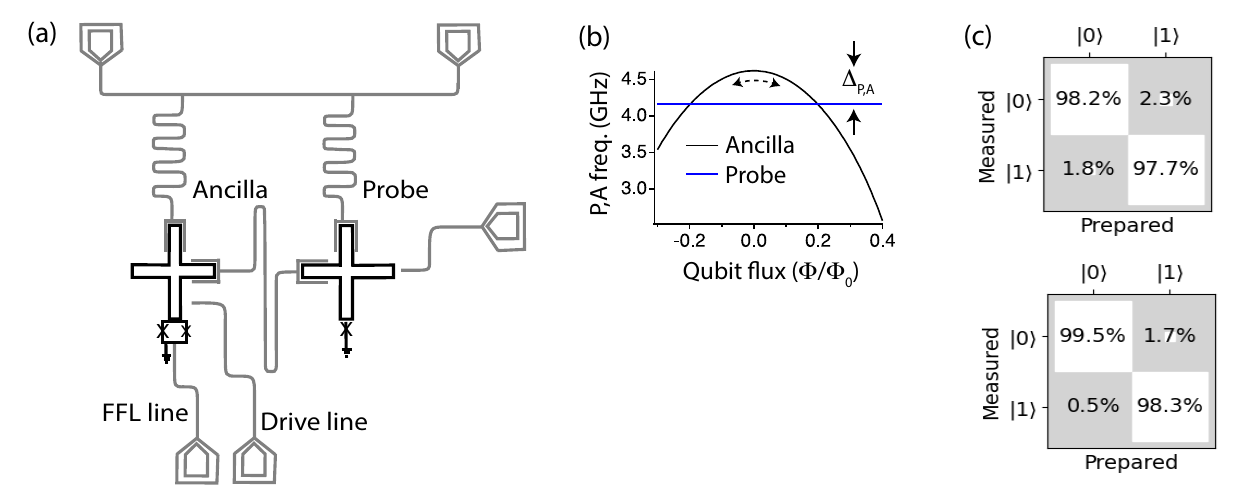}
    \caption{
        {\bf Experimental setup.} 
        (a) The experiment involves
        a two-qubit section of a transmon-circuit-based processor.
        (b) Parametric entangling gates are activated by modulation of the ancilla frequency via the fast-flux (FFL) line.
        (c) Readout-fidelity matrix for the probe (top) and ancilla (bottom).
        }
    \label{fig:setup}
\end{figure}

\begin{table}
\begin{center}
\begin{tabular}{||c | c | c |c|c|c|c|c||} 
\hline 
 & \thead{$\omega_\mathrm{q}/2\pi$ (GHz)} & \thead{$|\alpha|/2\pi$ (MHz)} & \thead{$\chi_\mathrm{qc}/2\pi$ (kHz)} &\thead{$\omega_\mathrm{c}/2\pi$ (GHz)} & \thead{$\kappa/2\pi$ (kHz)} & \thead{$T_1\ (\mu\mathrm{s})$} &\thead{$T_2^*\ (\mu\mathrm{s})$} \\ [0.5ex] 
\hline\hline
\thead{Ancilla qubit} & 4.2 & 212 &230&6.94 & 270 &32&41 \\ [0.5ex] 
\hline
\thead{Probe qubit} & 4.65 & 180 & 250& 7.09 & 206 &31&39\\ [0.5ex] 
 \hline
\end{tabular}
\end{center}
\caption{\textbf{Measured parameters of the device used in the experiment.}}
\label{tab:sim}
\end{table}

\end{itemize}

\subsection{Parametric gates} \label{sec:parametric}
   
In our setup, parametric gates have been found to achieve the highest-fidelity entangled states.   To implement such a gate, we modulate the ancilla qubit's frequency at approximately half of the probe--ancilla detuning, bringing the qubits into parametric resonance [Fig.~\ref{fig:setup}(b)].   When calibrated, the parametric resonance gives a probe--ancilla coupling rate of 0.954 MHz, corresponding to an $\sqrt{i\mathrm{SWAP}}$-gate time of 524 ns. We have optimized over the pulse envelope, obtaininga flux-modulation pulse with a gate time of 640 ns. We further find that entangling gate's parameters drift over time. Consequently, we stabilize the gate via feedback control. 

The gate's fidelity is estimated with quantum state tomography. We measure $9$ Pauli-expectation-value pairs $\langle \Sigma_\mathrm{P} \Sigma_\mathrm{A}\rangle$, with $\Sigma_\mathrm{P,A} \in \{X,Y,Z\}$, by measuring the qubits' reduced states simultaneously \cite{Kundu2019}. 
We use maximum-likelihood estimation 
\cite{James2001} to identify the components of the probe--ancilla density matrix.   

During the sensing protocol depicted in main text Fig.~3, we implement one $\sqrt{i\mathrm{SWAP}}$ gate to initialize the entangled state. Then, we projectively measure whether the system is in a 
singlet, $\ket{\Psi^-}$. That is, we measure $\{ \Pi_0, \mathbb{1} - \Pi_0 \}$ 
(as in the main text, $\Pi_0 \coloneqq \ketbra{ \Psi^- }{ \Psi^- }$).
This measurement consists of applying a second $\sqrt{i\mathrm{SWAP}}$ gate, which maps the singlet state to 
$\ket{e}_\mathrm{P}\ket{g}_\mathrm{A}$.
Because the qubits have different energies,
they accumulate different dynamical phases 
between the two $\sqrt{i\mathrm{SWAP}}$ gates. To remove this phase difference, we apply an additional physical rotation to the ancilla before the second $\sqrt{i\mathrm{SWAP}}$ gate.  We implement this phase correction with the physical rotation (\ref{Eq_RZ_as_R_X}) described above. 

\subsection{Effect of the parametric gate's fidelity on the QFI measurement}

As discussed in Sec.~\ref{sec:parametric}, we use a $\sqrt{i\mathrm{SWAP}}$ gate  to prepare and measure the singlet. We use quantum state tomography to determine the estimated density operator $\rho_\mathrm{exp}$. We then define the fidelity to a target state $\sigma = \ketbra{\Psi_-}{\Psi_-}$ as \cite{Nielsen11}
\begin{equation}
    \mathcal{F} = \Bigl(\mathrm{tr}\sqrt{\sqrt{\rho}\sigma\sqrt{\rho}}\Bigr)^2.
\end{equation}
Experimentally, we obtain $\mathcal{F} = 0.94(.02)$. To model the finite fidelity's effect on the measurement, we 
model the experimentally realized state as
\begin{equation} \label{eq:low_f_rho}
 \rho_\mathrm{exp} = \mathcal{F} \ket{\Psi_-}\bra{\Psi_-} + \frac{1}{3} (1-\mathcal{F}) (\mathbb{1} - \ket{\Psi_-}\bra{\Psi_-}).
\end{equation}
This is a mixture of the target state and a maximally mixed state.
The partial mixing reduces both the QFI that can be obtained and FI that we measure. 
From Eq.~\eqref{eq:low_f_rho}, we can calculate the FI about $\alpha$ by 
rotating the state, measuring the 
whether the system is in a singlet, repeating this process many times, and inferring the possible outcomes' probabilities. 
The resulting FI is
\begin{equation}\label{eq:finite_fidelity_FI}
   I_\alpha= - \frac{(1 - 4\mathcal{F})^2 \sin^2(\alpha)}{[-5 + 2\mathcal{F} + (-1 + 4\mathcal{F}) \cos(\alpha)][1 + 2\mathcal{F} + (-1 + 4\mathcal{F}) \cos(\alpha)]} \, .
\end{equation}

\begin{figure}
    \centering
    \includegraphics[width=8.0cm]{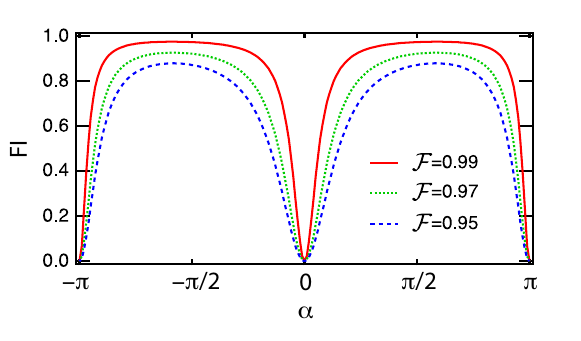}
    \caption{
        {\bf FI of the agnostic-sensing protocol versus singlet-preparation fidelity.} \kater{The curves are calculated from Eq.~\eqref{eq:finite_fidelity_FI}.}}
    \label{fig:finitefid}
\end{figure}

Figure \ref{fig:finitefid} displays the $I_\alpha$, calculated via Eq.~(\ref{eq:finite_fidelity_FI}), versus $\alpha$. We note two important trends. First, the FI now depends on $\alpha$, indicating that the singlet--triplet measurement's metrological performance depends on the rotation's magnitude. 
Second, the imperfect singlet-preparation fidelity crucially limits the advantage of manipulating entanglement 
to perform rotation-axis-agnostic sensing.  
A general analysis of the limits on the FI, in the presence of finite-fidelity operations, will be an intriguing and important direction for future research.

\section{Theoretical analysis of sensing protocols}

Here, we discuss theoretical background for the four sensing protocols laid out in the main text. 
First, we review the estimation of phases from locally unbiased estimators (App.~\ref{sec:localestimation}). In App.~\ref{App_Agnostic_Phase_Estimation}, we identify a single-qubit sensor's limitations. A qubit has only two DOFs, whereas three DOFs specify an arbitrary rotation. Hence more than one qubit is necessary for sensing a rotation about an unknown axis.
In App.~\ref{Appendix_Single_Qubit}, we detail the entanglement-free sensor whose qubit probe is tagged by an ancilla. 
A similar strategy underlies quantum process tomography~\cite{Chuang97, DAriano01, Altepeter03, Song2021}.
This sensor's QFI is 2/3, we show.
In App.~\ref{Appendix_Bell_Sensing}, we prove that the qubit probe entangled with an ancilla achieves the maximum QFI over input states, 1. 
Table~\ref{tab:thy} summarizes (i) the resources required by each sensing strategy and (ii) the strategy's effectiveness, as quantified by the QFI.

\begin{table}
\begin{center}
\begin{tabular}{||l | c ||} 
\hline 
  \thead{Resource} & QFI, \thead{$ \mathcal{I}_\alpha$} \\ 
\hline\hline
\thead{Qubit probe} & --  \\
\hline
\thead{Qubit probe \& classical ancilla} & 2/3 \\ [0.5ex] 
 \hline
 \thead{Qubit probe \& entangled ancilla } & 1  \\ [0.5ex] 
 \hline
\end{tabular}
\end{center}
\caption{ \textbf{Summary of theoretical results.}
}
\label{tab:thy}
\end{table}

\subsection{Summary of local estimation theory} \label{sec:localestimation}

A typical quantum estimation problem is structured as follows.
$M$ continuous parameters $\bm{\alpha} \coloneqq (\alpha_1, \alpha_2 , \dots, \alpha_M)$ are encoded in a parameterized quantum state $\rho_{\bm{\alpha}}$. The goal is to infer the parameters' values. 
This parameter-estimation problem 
is addressed within the field of quantum estimation theory.
We focus on the well-established subfield of local estimation, in which $\bm{\alpha}$ is unknown but fixed~\cite{paris2009quantum}. One aims to minimize the estimators' covariances, at fixed values of the parameters.
A general scheme consists of three steps~\cite{Giovanetti11}:
\begin{enumerate}
    \item Prepare a parameter-independent probe state $\rho_0$. 
    \item Evolve the state under a parameter-dependent unitary $U_{\bm{\alpha}}$:
    $\rho_{\bm{\alpha}} = U_{\bm{\alpha}} \rho_0 U_{\bm{\alpha}}^{\dagger} \, .$
    \item Measure the state. The most general measurement is a positive-operator-valued measure (POVM) \cite{Giovannetti06}. A POVM is a set $\big\{ {F}_k\big\}$ of positive-semidefinite operators $({F}_k \geq 0)$ that sum to the identity ($\sum_k {F}_k = {\mathbb{1}}$). The measurement yields outcome $k$ with a probability
    $p(k|\bm{\alpha}) = \text{Tr} \big[ {F}_{k} \rho_{\bm{\alpha}} \big] \, .$  
\end{enumerate}
The probability distribution $ p(k|\bm{\alpha})$ encapsulates information about the parameters $\alpha_j$. 
 
The parameters are estimated through an estimator
$
\hat{\bm{\alpha}}(k)\equiv \hat{\bm{\alpha}}   \coloneqq (\hat{\alpha}_1 , \hat{\alpha}_2 , \dots \hat{\alpha}_M  \big)$. 
Each $\hat{\alpha}_i$ implicitly has an argument $k$, which we sometimes omit for conciseness.
An \emph{estimator} is a map from the space of measurement outcomes to the space of possible parameter values. 
We use 
\emph{locally unbiased} estimators $\hat{\bm{\alpha}}$:
\begin{equation} \label{Eq_Unbiased_Constraints}
     \sum_{k}{\big[ \bm{\alpha} - \hat{\bm{\alpha}}(k) \big]   \, p(k|\bm{\alpha}) } =0 
     \, , \; \: \text{and} \; \:
     \sum_{k}  {\hat{\alpha}_i(k) \,  
     \partial_j p(k|\bm{\alpha}) } 
     = \delta_{i j }  \, .
\end{equation}
The indices $i,j \in \{ 1, 2, \dots , M \}$, and $\partial_j \coloneqq \partial/\partial \alpha_j$.
The constraints in Eq.~\eqref{Eq_Unbiased_Constraints}  ensure that the estimator tracks the parameter's true value faithfully, to first order around the point $\bm{\alpha}$. The second constraint excludes pathological estimators. Examples of such include estimators that return a fixed value, irrespective of the measurement outcome. 

For unbiased estimators, the accuracy of $\hat{\bm{\alpha}}$ is quantified by its covariance matrix, 
\begin{equation} \label{Eq_Covariance_Def}
    \text{Cov} \big( \hat{\bm{\alpha}} \big) 
    \coloneqq \sum_{k}{\big[ \bm{\alpha} - \hat{\bm{\alpha}}(k) \big] \big[ \bm{\alpha} - \hat{\bm{\alpha}}(k) \big]^{\top} p(k|\bm{\alpha}) } \, .
\end{equation}
Using Eqs. \eqref{Eq_Unbiased_Constraints} and \eqref{Eq_Covariance_Def}, 
one can show that the covariance matrix obeys the Cram\'er--Rao bound \cite{Cramer1946, Rao1992},
\begin{equation} \label{Eq_1}
   \text{Cov} \big( \hat{\bm{\alpha}} \big) 
   \geq \frac{1}{N} \, I(\bm{\alpha})^{-1} \, .
\end{equation}
$N$ denotes the number of measurements, and $I(\bm{\alpha})$ denotes the \emph{Fisher-information matrix} (FIM):
\begin{equation}
    I (\bm{\alpha})_{i,j} \coloneqq \sum_{k}{p(k|\bm{\alpha}) \big[ \partial_i \log{p(k|\bm{\alpha})} \big] \big[ \partial_j \log{p(k|\bm{\alpha})} \big]} \, .
\end{equation}
The FIM relates how easily we can distinguish neighboring probability distributions (parameterized by close-together $\bm{\alpha}$ values).
Therefore, the FIM quantifies the information that a probability distribution encodes about the parameters $\alpha_i$.
The choice of measurement (step 3) affects the probabilities $p(k|\bm{\alpha})$ and hence $I (\bm{\alpha})$. 
Certain measurements maximize the amount of information extractable form the probe.
For every measurement, the inverse FIM is lower-bounded  by the inverse \textit{quantum}-Fisher-information matrix (QFIM):
\begin{equation}
    I(\bm{\alpha})^{-1} \geq \mathcal{I} (\bm{\alpha} | \rho_{\bm{\alpha}})^{-1} \, . \label{Eq_QFIM_CFIM}
\end{equation}
That is, $I(\bm{\alpha})^{-1} - \mathcal{I} (\bm{\alpha} | \rho_{\bm{\alpha}})^{-1}$ is positive-semidefinite. The QFIM is defined as~\cite{Braunstein94, Fujiwara1995, Liu2019} 
\begin{equation} \label{Eq_Def_QFIM}
    \mathcal{I}(\bm{\alpha}|\rho_{\bm{\alpha}})_{i,j}  \coloneqq\text{Tr}\big[ \Lambda_{i} \, \partial_j  \rho_{\bm{\alpha}}  \big] \, ,
\end{equation}
where $\Lambda_{i}$ denotes the \emph{symmetric logarithmic derivative} (SLD),
defined implicitly by 
$\partial_{i} \rho_{\bm{\alpha}} 
= \dfrac{1}{2} (\Lambda_{i} \rho_{\bm{\alpha}} + \rho_{\bm{\alpha}} \Lambda_{i} )$ \citep{Helstrom1969}. 

By the bound~\eqref{Eq_QFIM_CFIM},   
the QFIM can replace the FIM in Eq.~\eqref{Eq_1}. The replacement leads to the \emph{quantum Cram\'er--Rao bound} (QCRB),
    \begin{equation}\label{eqn:QCR_bound}
     \text{Cov} \big(  \hat{\bm{\alpha}} \big) 
     \geq \frac{1}{N} \, \mathcal{I} (\bm{\alpha}| \rho_{\bm{\alpha}})^{-1} \, .
    \end{equation}
Before analyzing the QCRB in full, we impart intuition through the special case of single-parameter estimation.

Consider estimating only one parameter, $\alpha$ ($M = 1$).
The matrix inequality~\eqref{eqn:QCR_bound} becomes a scalar inequality. Furthermore, in the limit $N \rightarrow \infty$,
the bound can always be saturated~\cite{Braunstein94}. To elucidate the saturation, we denote by $\Lambda_{\alpha}$ the SLD operator associated with the parameter $\alpha$. Consider a POVM whose elements project onto the eigenspaces of $\Lambda_{\alpha}$. This POVM saturates the QCRB.

Now, consider estimating multiple parameters. The matrix inequality~\eqref{eqn:QCR_bound} is not generally saturable, for the following reason. An optimal measurement basis, from which to infer about $\alpha_i$, is the eigenbasis of $\Lambda_i$. This measurement basis may be far from optimal for inferring about $\alpha_j$.
That is, parameters' SLD operators $\Lambda_{i}$ might not (weakly) commute with each other. 
If they do not, then one cannot achieve the optimal precision for all the parameters simultaneously~\cite{zhu2015, heinosaari2016, ragy2016,Albarelli2020}.

Furthermore, the bound~\eqref{eqn:QCR_bound} has another problem, even in classical estimation theory. Consider two experimental strategies specified by two distinct POVMs, $\Pi_1$ and $\Pi_2$. Let $I_1(\bm{\alpha})$ and $I_1(\bm{\alpha})$ denote the corresponding FIMs. There can be situations in which both $I_1(\bm{\alpha})^{-1} \nleq I_2(\bm{\alpha})^{-1}$ and $I_2(\bm{\alpha})^{-1} \nleq I_1(\bm{\alpha})^{-1}$. 
The covariance matrices $\text{Cov}_1 \big(  \hat{\bm{\alpha}} \big)$ and $\text{Cov}_2 \big(  \hat{\bm{\alpha}} \big)$ lack partial ordering, defined in terms of the positive-semidefinite relation. Therefore, which strategy performs best can be unclear.

To compare strategies effectively, we introduce a scalar bound, using a real, positive-semidefinite $M \times M$ \emph{weight matrix} $W$. In terms of $W$, the matrix bound ~\eqref{eqn:QCR_bound} is recast as
\begin{equation} \label{Eq_Scalar_Risk_Function}
    \text{Tr}[ W \text{Cov} \big( \hat{\bm{\alpha}} \big)] \geq  \frac{1}{N} \text{Tr}[ W \mathcal{I} (\bm{\alpha} | \rho_{\bm{\alpha}})^{-1}  ]  \, .
\end{equation}
We choose $W$ in accordance with the parameter-estimation experiment's goal. In the next subsection (\ref{App_Agnostic_Phase_Estimation}), we choose a weight matrix to use throughout the remainder of the appendix. Particular choices of $W$ can lead to 
different optimal estimation strategies ~\cite{Goldberg2021}.
For instance, let $W = \mathbb{1}$, and let $\hat{\bm{\alpha}}$ be an arbitrary unbiased estimator. The left-hand side of Eq.~\eqref{Eq_Scalar_Risk_Function} equals the sum of the mean-square errors of the parameters in $\hat{\bm{\alpha}}$.

Inequality~\eqref{Eq_Scalar_Risk_Function} motivates the QFIM as a useful performance metric in multiparameter metrology. However, for Ineq.~\eqref{Eq_Scalar_Risk_Function} is not generally saturable. 
However, for locally unbiased estimators, in the asymptotic limit of many measurements (as $N \rightarrow \infty$) ~\cite{Albarelli2020}, 
\begin{equation} \label{Eq_Scalar_Risk_Function_2}
   \frac{1}{N} \text{Tr}[ W \mathcal{I} (\bm{\alpha} | \rho_{\bm{\alpha}})^{-1}  ] \leq \underset{\mathcal{M}}{\text{min}} \text{Tr}[ W \text{Cov} \big( \hat{\bm{\alpha}} \big)] \leq 2 \frac{1}{N} \text{Tr}[ W \mathcal{I} (\bm{\alpha} | \rho_{\bm{\alpha}})^{-1}  ]  \, .
\end{equation}
where $\mathcal{M}$ is the set of all possible measurements. Therefore, the QFIM bounds the covariance to within a factor of 2. 

Often, one repeats the protocol $N>1$ times, using identical copies of the state $\rho_0$.
One measures 
$ \rho_{\bm{\alpha}}^{\otimes N}$. 
The QFIM turns out to be additive,
$\mathcal{I} (\bm{\alpha}| \rho_{\bm{\alpha}}^{\otimes N})= N \, \mathcal{I} (\bm{\alpha}| \rho_{\bm{\alpha}})$, implying that 
\begin{equation} \label{Eq_Scalar_Risk_Function_N}
    \text{Tr} \left[ W \text{Cov} \big( \hat{\bm{\alpha}} \big) \right]  \geq \text{Tr} \left[ W \mathcal{I} (\bm{\alpha}| \rho_{\bm{\alpha}}^{\otimes N})^{-1}  \right]
    = \dfrac{1}{N} \text{Tr} \left[ W \mathcal{I} (\bm{\alpha} | \rho_{\bm{\alpha}})^{-1}  \right]  \, .
\end{equation}
Hence, one can decrease the estimator's variance in two ways: first, one can increase the number $N$ of measurements.
Second, one can design a setup that reduces $\text{Tr}[ W \mathcal{I} (\bm{\alpha} | \rho_{\bm{\alpha}})^{-1}  ]$. 

Finally, 
one can optimize the $\rho_0$ in step 1. The QFIM $\mathcal{I} (\bm{\alpha}|\rho_{\bm{\alpha}})$ is convex \cite{Fujiwara2001}. Therefore, pure probe states achieve the maximum QFIM~\cite{Pang2014, Pang2015}.

\subsection{Agnostic phase estimation} \label{App_Agnostic_Phase_Estimation}

In the experiments described in the main text, the probe undergoes an arbitrary unknown rotation, represented by $U_{\bm{\alpha}} = \exp \left(-i \alpha \hat{\bm{n}}\cdot\bm{\sigma} / 2 \right)$. 
The unknown rotation angle is $\alpha$, 
$\hat{\bm{n}} = \sin \theta \cos \phi \, \hat{\bm{x}} 
+ \sin \theta \sin \phi \, \hat{\bm{y}} 
+ \cos \theta \, \hat{\bm{z}}$ defines the unknown rotation axis, and $\bm{\sigma} = (X, Y, Z)$ denotes a vector of Pauli operators.
The number of unknown parameters is $M=3$. 
One unknown parameter is $\alpha$. The other two, $\theta$ and $\phi$,  parameterize $\hat{\bm{n}}$. In the rest of this appendix, we will denote by $\bm{\alpha} \coloneqq (\alpha, \theta, \phi)$ the parameters to be estimated.

We aim to calculate the precision with which $\alpha$ can be estimated. We do not aim to estimate $\theta$ and $\phi$.
Hence, we set the weight matrix to
\begin{equation}
  W = 
  \begin{pmatrix}
1 & 0 & 0 \\
0 & 0 & 0 \\
0 & 0 & 0 \\
\end{pmatrix} \, .
\end{equation}
The scalar bound~\eqref{Eq_Scalar_Risk_Function_N}
becomes
\begin{equation} \label{Eq_Specific_Bound}
    \text{Var} \big( \hat{\alpha} \big) \geq \dfrac{1}{N} \,  {\mathcal{I} ( \bm{\alpha} | \rho_{\bm{\alpha}} )^{-1}}_{1,1} \, . 
\end{equation}
Due to the form of $W$, if the QFIM takes the form
\begin{equation}
\label{Eq_Special_QFIM}
    \mathcal{I} ( \bm{\alpha} | \rho_{\bm{\alpha}}) =   \begin{pmatrix}
\mathcal{I}_{\alpha}  & 0 & 0 \\
0 & \star & \star \\
0 & \star & \star \\
\end{pmatrix}  \; \; \text{for all } \; \; \bm{\alpha} ,
\end{equation}
then the variance $\text{Var} \big( \hat{\alpha} \big) \geq 1/(N \mathcal{I}_{\alpha} ) $, irrespectively of the lower right-hand $2 \times 2$ block $\mathcal{I} ( \bm{\alpha} | \rho_{\bm{\alpha}})$. 
(Multiplication by $W$ always maps this block's inverse to zero.)
We denote the QFI of $\alpha$ by $ \mathcal{I}_{\alpha}$, to distinguish it from the QFIM, $\mathcal{I} (\bm{\alpha}| \rho_{\bm{\alpha}})$. When the QFIM is singular and not in the form~\eqref{Eq_Special_QFIM}, one cannot estimate $\alpha$ without knowledge of $\hat{\bm{n}}$.

In the next section, we will see that, for every pure initial state, the single-qubit-probe protocol leads to a singular 
$\mathcal{I} ( \bm{\alpha} | \rho_{\bm{\alpha}} )$, which does not decompose in the form~\eqref{Eq_Special_QFIM}. 
Therefore, this protocol cannot produce any estimate of $\alpha$. 
In contrast, we show, ancilla-assisted entanglement-free sensing and entangled-sensor sensing lead to QFIMs of the desired form~\eqref{Eq_Special_QFIM}. The QFIs for $\alpha$ are $2/3$ and $1$, respectively. Hence entanglement can boost the QFI of $\alpha$ by $50\  \%$.

\subsection{Single-qubit sensor initiated in a pure state} 
\label{Appendix_Single_Qubit}

We first address a particular pure state, then generalize.
Consider preparing a single-qubit probe in
$\ket{\psi_0} = \ket{0}$, the  eigenvalue-1 eigenstate of $Z$. The probe undergoes an arbitrary unknown rotation, represented by $U_{\bm{\alpha}} = \exp \left(-i\alpha\hat{\bm{n}}\cdot\bm{\sigma} / 2 \right)$, as detailed in the previous subsection. The state $\ket{\psi_{\bm{\alpha}}} = U_{\bm{\alpha}} \ket{0}$ achieves the QFIM
\begin{equation} \label{Eq_QFIM_Sz}
    \begin{pmatrix}
\sin^2(\theta) & -2 \cos(\theta) \sin^2(\theta) \sin^2\left(\dfrac{\alpha}{2}\right) & \sin(\alpha)  \cos(\theta) \sin(\theta)\\
    -2 \cos(\theta) \sin^2(\theta) \sin^2\left(\dfrac{\alpha}{2}\right) & 4 \sin^2(\theta) \sin^2\left(\dfrac{\alpha}{2}\right)\left(1 - \sin^2(\theta) \sin^2\left(\dfrac{\alpha}{2}\right)\right) & 2 \sin^3(\theta) \sin^2\left(\dfrac{\alpha}{2}\right) \sin(\alpha) \\
     \sin(\alpha)  \cos(\theta) \sin(\theta) & 2 \sin^3(\theta) \sin^2\left(\dfrac{\alpha}{2}\right) \sin(\alpha) & 4 \sin^2\left(\dfrac{\alpha}{2}\right) \left(\cos^2(\theta) + \sin^2(\theta) \sin^2\left(\dfrac{\alpha}{2}\right)\right)
\end{pmatrix} \, . 
\end{equation}
This matrix's determinant is always zero; the matrix is not invertible. Indeed, the first line of Eq.~\eqref{Eq_QFIM_Sz} is a linear combination of the second and third lines. We can check this fact by adding the second line, multiplied by $-\frac{1}{2} \sec{\theta}$, to the third line,  multiplied by $\frac{1}{2} \cot{\left(\alpha/2 \right)} \tan{\theta}$.
Furthermore, for most $\bm{\alpha}$ values, the QFIM does not satisfy Eq.~\eqref{Eq_Special_QFIM}.

Similarly, we can calculate the QFIM for any other pure input state $\ket{\psi_0} = a \ket{0} + b\ket{1}$, with $a, b \in \mathcal{C}$ and $|a|^2 + |b|^2 = 1$. The evolved state is $\ket{\psi_{\bm{\alpha}}} = U_{\bm{\alpha}} \ket{\psi_0}$.  The corresponding QFIM is not invertible, and there are values of $\bm{\alpha}$ for which 
the QFIM does not satisfy Eq.~\eqref{Eq_Special_QFIM}.  
We conclude that $\alpha$  cannot  be estimated from a qubit probe initiated in any pure state.

We can understand this conclusion intuitively through an example. Consider aiming to measure the strength of a magnetic field $\bm{B}$ whose magnitude and direction are unknown. Suppose that we employ three (non-coplanar) detectors, each measuring the field component that points along some axis (e.g., $B_x$, $B_y$, or $B_z$).
The magnitude of $\bm{B}$ (or $\alpha$, in our case) follows from $|\bm{B}| = \sqrt{B_x ^2 + B_y ^2 + B_z ^2}$. 
The field's direction (or $\hat{\bm{n}}$, in our case) follows from $\cos{\phi} = B_x / \sqrt{B_x ^2 + B_y ^2}$ and  $\cos{\theta} = B_z / |\bm{B}|$. Three parameters specify the field, so one cannot measure the field's magnitude and direction using only one detector (or, in our case, a qubit probe in a pure state).

\subsection{Ancilla-assisted entanglement-free sensing}
\label{Appendix_Ancilla_Classical}

We can enhance the pure-state single-qubit probe (App.~\ref{Appendix_Single_Qubit}) even without introducing entanglement. Before showing how to do so, we elucidate a subtlety of the term \emph{single-qubit probe.} Each of our sensing strategies requires many identical trials. Hence even the single-qubit-probe strategy (App.~\ref{Appendix_Single_Qubit}) requires many qubit probes. However, those qubits are prepared and used independently of each other. 

This subsection likewise concerns a sensing strategy that involves many identical trials. In each trial, a single-qubit probe is prepared in a pure state $\ket{\psi_j}$ selected randomly according to a probability distribution $\{ p_j \}$.
An ancilla in the state $\ket{j}$ records which $\ket{\psi_j}$ was prepared. The ancilla may be a qubit or a classical bit. In  experiments, one tends to keep a classical record of which $\ket{\psi_j}$ was prepared.
The probe and ancilla begin in the mixture
\begin{equation} 
   \label{Eq_DensityMatrix_Ancilla_Classical}
   \rho_0 = \sum_j {p_j \ketbra{\psi_j}{\psi_j} \otimes \ketbra{j}{j}} \, , \; \text{wherein} \; 
\sum_j p_j = 1 .
\end{equation}
Suppose that one lacks knowledge of $\alpha$ and $\hat{\bm{n}}$. 
The prior distribution most reasonably attributable to $\hat{\bm{n}}$ is uniform. We define an \emph{optimal} entanglement-free input state as a state that achieves the first two equalities in Eq.~2, main text, on average with respect to the uniform distribution over $\hat{\bm{n}}$.
We will show that the optimal input states have the form
\begin{align} \label{Eq_General_Optimal_Classical_Ancilla_1}
    \rho_0 & = \dfrac{1}{3} \Big( \ketbra{z \pm}{z \pm} \otimes \ketbra{0}{0} + e^{i \phi_x}  \ketbra{x \pm}{x \pm} \otimes \ketbra{1}{1} 
    +  e^{i \phi_y} \ketbra{y \pm}{y \pm} \otimes \ketbra{2}{2}\Big) \, .
\end{align}
The bases $\{ \ket{x \pm} \}$, $\{ \ket{y \pm} \}$, and $\{ \ket{z \pm} \}$ (as defined in the main text) are \emph{mutually unbiased}: for every 
$a, b \in \{x,y,z\}$ such that $a \neq b$, 
$| \braket{a \pm}{b \pm} | 
= | \braket{a \pm}{b \mp} | 
= 1 / \sqrt{d}$, wherein $d=2$ denotes the Hilbert space's dimensionality.
$ \rho_0$ is a $1:1:1$ mixture of elements of three mutually maximally incompatible bases. The relative phases $e^{i \phi_x}$ and $e^{i \phi_y}$ are arbitrary.

In the experimental implementation, we program three different pulse sequences. Each sequence prepares one pure state ($\ket{x\pm}$, $\ket{y \pm}$, or $\ket{z \pm}$), 
applies the unknown unitary, and then measures the probe in a chosen basis. For each sequence, a register stores the ancilla label 
(effectively, 0, 1, or 2), 
which we use to determine which measurement (mentioned above) to perform.
We perform enough experimental trials to reduce the binomial errors in our estimates of the measurement outcomes' probabilities.

Now, we show that the family of states~\eqref{Eq_General_Optimal_Classical_Ancilla_1} is  optimal. First, we bound the best precision attainable with any state of the form~\eqref{Eq_DensityMatrix_Ancilla_Classical},
averaged over all directions $\hat{\bm{n}}$. Next, we show that the state~\eqref{Eq_General_Optimal_Classical_Ancilla_1}
saturates this bound. This state leads to a QFIM of the form~\eqref{Eq_Special_QFIM}, with a QFI for $\alpha$ of $\mathcal{I}_{\alpha} = 2/3$. Recall that the entanglement-assisted-sensing strategy (discussed in the main text) leads to $\mathcal{I}_{\alpha} = 1$. The entanglement-assisted strategy therefore performs $50$ \% better than the best entanglement-free strategy. 

\subsubsection{Bound on the best precision attainable with ancilla-assisted entanglement-free sensing} \label{Appendix_Bound_Ent_Free}

In this section, we bound the best precision attainable without entanglement. Equation~\eqref{Eq_DensityMatrix_Ancilla_Classical} specifies the input state.
The unitary $U_{\bm{\alpha}} = \exp \left(-i\alpha\hat{\bm{n}}\cdot\bm{\sigma} / 2 \right)$ evolves the probe state to
\begin{align}
    \rho_{\bm{\alpha}} 
    & = \sum_j {p_j \, U_{\bm{\alpha}} \ketbra{\psi_j}{\psi_j} U_{\bm{\alpha}}^{\dagger}  \otimes \ketbra{j}{j}} 
    \equiv \sum_j {p_j \, \rho_{\bm{\alpha}}^{(j)}  \otimes \ketbra{j}{j}} \, ,
\end{align}
where $\rho_{\bm{\alpha}}^{(j)} 
\coloneqq U_{\bm{\alpha}} \ketbra{\psi_j}{\psi_j} U_{\bm{\alpha}}^{\dagger} $. 
We calculate the QFIM achievable with $ \rho_{\bm{\alpha}}$, using the (extended) convexity bound on the QFIM~\cite{Alipour2015}, 
\begin{equation} \label{Eq_Convexity_Bound_Ineq}
    \mathcal{I} ( \bm{\alpha}| \rho_{\bm{\alpha}} ) 
    \leq I ( \bm{\alpha} | \{p_j \} )
    + \sum_{j=1}^N p_j \,  
    \mathcal{I} \left( \bm{\alpha} |  \rho_{\bm{\alpha}}^{(j)}  \otimes \ketbra{j}{j} \right) \, .
\end{equation}
$I ( \bm{\alpha} | \{p_j \} )$ denotes the 
FIM, which
describes the information accessible via sampling from the probability distribution $\{ p_j  \}$.
In this case, the convexity bound is saturated \cite{Wilfred2024}. The reason is that the states $ \rho_{\bm{\alpha}}^{(j)}  \otimes \ketbra{j}{j}$ are mutually orthogonal.
\begin{equation}
    \mathcal{I} ( \bm{\alpha}| \rho_{\bm{\alpha}} ) 
    = I ( \bm{\alpha} | \{p_j \} )
    + \sum_{j=1}^N p_j \,  
    \mathcal{I} \left( \bm{\alpha} |  \rho_{\bm{\alpha}}^{(j)} \otimes \ketbra{j}{j} \right) \, .
\end{equation}
The probabilities $p_j$ depend on neither $\alpha$ nor $\hat{\bm{n}}$. The first term therefore vanishes: 
$I( \bm{\alpha} | \{p_j \}) = 0 $. 
Furthermore, the QFIM is additive, since the ancilla state $\ketbra{j}{j}$ is parameter-independent: 
$\mathcal{I} \left( \bm{\alpha} |  \rho_{\bm{\alpha}}^{(j)} \otimes \ketbra{j}{j} \right) 
 = \mathcal{I} \left( \bm{\alpha} |  \rho_{\bm{\alpha}}^{(j)} \right) + \mathcal{I} \left( \bm{\alpha} |  \ketbra{j}{j} \right) = \mathcal{I} \left( \bm{\alpha} |  \rho_{\bm{\alpha}}^{(j)}  \right)$. Therefore,
\begin{equation} 
   \label{Eq_Convexity_Simplified}
    \mathcal{I}(\bm{\alpha} | \rho_{\bm{\alpha}} ) 
    =  \sum_{j=1}^N p_j \, \mathcal{I}(\bm{\alpha} | \rho_{\bm{\alpha}}^{(j)}) \, .
\end{equation}

Recall that we aim to lower-bound the variance of $\alpha$, using Eq.~\eqref{Eq_Specific_Bound}. We need the $(1,1)$ component of the inverse QFIM, ${\mathcal{I}(\bm{\alpha} | \rho_{\bm{\alpha}} ) ^{-1}}_{1,1}$. 
In App.~\ref{Appendix_Single_Qubit}, we showed that the matrices $\mathcal{I}(\bm{\alpha} | \rho_{\bm{\alpha}})$ are neither invertible nor generally diagonal. 
Therefore, calculating the inverse QFIM, $\mathcal{I}(\bm{\alpha} | \rho_{\bm{\alpha}} )^{-1}$, is nontrivial. Instead, we provide a  method for lower-bounding ${\mathcal{I}(\bm{\alpha} | \rho_{\bm{\alpha}} ) ^{-1}}_{1,1}$. 

The method begins with the following expression for the $3 \times 3$ QFIM for $\bm{\alpha}$:
\begin{equation}
    \mathcal{I}(\bm{\alpha} | \rho_{\bm{\alpha}} ) =
    \begin{pmatrix}
        \mathcal{I}_{\alpha} & \mathcal{I}_{\text{c}} ^{T} \\
        \mathcal{I}_{\text{c}} &\mathcal{I}_{\hat{\bm{n}}}
    \end{pmatrix}  \, .
\end{equation}
$\mathcal{I}_{\text{c}}$ denotes a two-column vector that quantifies correlations between $\alpha$ and $\hat{\bm{n}}$. $\mathcal{I}_{\hat{\bm{n}}}$ is a $2\times 2$ matrix that quantifies the information about $\hat{\bm{n}}$. 
If $\mathcal{I}(\bm{\alpha} | \rho_{\bm{\alpha}} )$ is invertible, one can represent its inverse using the Schur complement~\cite{Gallier2010}:
\begin{equation}
     \mathcal{I}(\bm{\alpha} | \rho_{\bm{\alpha}} ) ^{-1} = 
    \begin{pmatrix}
        \dfrac{1}{\mathcal{I}_{\alpha} -\mathcal{I}_{\text{c}} ^{T} \, \mathcal{I}_{\hat{\bm{n}}} \, \mathcal{I}_{\text{c}}} & *  \\
        * & *
    \end{pmatrix} \, .
\end{equation}
Since $\mathcal{I}(\bm{\alpha} | \rho_{\bm{\alpha}} )$ is 
positive-semidefinite, also $\mathcal{I}_{\hat{\bm{n}}}$ is positive-semidefinite. (A positive-semidefinite
$d \times d$ matrix $A$ satisfies $x^T A x \geq 0$ for all $x$ in $\mathbb{R} ^ d$.) All the QFIM's entries are real, so the column vector $\mathcal{I}_{\text{c}}$ is in $\mathbb{R} ^ 2$.  Hence, $\mathcal{I}_{\text{c}} ^{T} \, \mathcal{I}_{\hat{\bm{n}}} \, \mathcal{I}_{\text{c}}$ is non-negative. Therefore, 
\begin{equation} \label{Eq_Inequality}
    \mathcal{I}_{\alpha} \geq \mathcal{I}_{\alpha} -\mathcal{I}_{\text{c}} ^{T} \, \mathcal{I}_{\hat{\bm{n}}} \, \mathcal{I}_{\text{c}}  \, .
\end{equation}
Inverting each side of the inequality yields
\begin{align}
   \dfrac{1}{\mathcal{I}_{\alpha} -\mathcal{I}_{\text{c}} ^{T} \, \mathcal{I}_{\hat{\bm{n}}} \, \mathcal{I}_{\text{c}} } \geq \dfrac{1}{\mathcal{I}_{\alpha} } 
   \quad \Leftrightarrow \quad
    {\mathcal{I}(\bm{\alpha} | \rho_{\bm{\alpha}} ) ^{-1}}_{1,1} \geq \dfrac{1}{{\mathcal{I}(\bm{\alpha} | \rho_{\bm{\alpha}} )}_{1,1}} \, .
    \label{eq_bound_help1}
\end{align}
By Eq.~\eqref{Eq_Convexity_Simplified}, the $(1,1)$ matrix element
$\mathcal{I}(\bm{\alpha} | \rho_{\bm{\alpha}} )_{1,1}$ equals a weighted sum of the elements $\mathcal{I}(\bm{\alpha} | {\rho_{\bm{\alpha}}}^{j} )_{1,1}$:
$\mathcal{I}(\bm{\alpha} | {\rho_{\bm{\alpha}}} )_{1,1} =  \sum_j p_j \, \mathcal{I}(\bm{\alpha} | \rho_{\bm{\alpha}}^{(j)})_{1,1} \, .$
Recall that the prior distribution most reasonably attributable to $\hat{\bm{n}}$ is uniform.
We must average over this distribution to evaluate the expected performance of our estimate of $\alpha$. 
We can substitute into the right-hand side of Ineq.~\eqref{eq_bound_help1}, if $\mathcal{I}(\bm{\alpha} | \rho  )$ is invertible:
\begin{align} \label{Eq_LowerBound}
    \text{Avg}_{\hat{\bm{n}}} \big[ {\mathcal{I}(\bm{\alpha} | \rho_{\bm{\alpha}} ) ^{-1}}_{1,1} \big]
    & \geq \dfrac{1}{\sum_j p_j  \text{Avg}_{\hat{\bm{n}}}\big[ \mathcal{I}(\bm{\alpha} | \rho_{\bm{\alpha}}^{(j)})_{1,1} \big] } 
    =\dfrac{1}{2/3 \sum_j p_j } 
    = \dfrac{3}{2} \, .
\end{align}
$\text{Avg}_{\hat{\bm{n}}}$ denotes the average with respect to a uniform distribution over the axes $\hat{\bm{n}}$. 
The first equality follows from a fact proved in the next subsubsection: consider preparing an arbitrary single-qubit pure state. It achieves a QFI that, averaged over all $\hat{\bm{n}}$, is $2/3$.
That is, $\text{Avg}_{\hat{\bm{n}}}\big[ \mathcal{I}(\bm{\alpha} | \rho_{\bm{\alpha}}^{(j)})_{1,1} \big] = 2/3$.

According to the argument above, every ancilla-assisted entanglement-free input state $\rho_0$ leads to an average-over-$\hat{\bm{n}}$ quantum Fisher information $\mathcal{I}_{\alpha} = 2/3$. For such an input to be optimal, it must achieve  the first two equalities in Eq.~2, main text. 
$\rho_0$ achieves this condition only if
the estimation of $\alpha$ is independent of the estimation of $\hat{\bm{n}}$. 
As discussed in Sec.~\ref{Appendix_Bound_Ent_Free}, this happens if and only if
$\mathcal{I}_{\hat{\bm{n}}} = 0$. 
When $\mathcal{I}_{\hat{\bm{n}}} = 0$, Ineq.~\eqref{Eq_Inequality} is saturated, and hence the bound~\eqref{Eq_LowerBound} is achieved. We therefore reach a necessary
criterion for an entanglement-free input state to be optimal: the state must entail a QFIM of the form ~\eqref{Eq_Special_QFIM}.

In Sec.~\ref{Sec_App_Entanglement_Free_Strategy}, we derive the optimal input states' form. 
We prove these states' optimality by computing the QFIM of a general one of these states.
We calculate the QFIM using a convexity bound on the QFIM 
and exploiting the fact that the convexity bound is saturated in this case~\cite{Wilfred2024}.

\subsubsection{Average quantum Fisher information achievable with arbitrary pure single-qubit state} \label{App_avg_QFI_n}

We now prove that, for every pure single-qubit state, averaging the QFI of $\alpha$ over all $\hat{\bm{n}}$ yields $2/3$.
Our proof strategy is direct calculation. First, we recall the forms of $\hat{\bm{n}}$, $A$ 
and $U_\alpha$ in terms of the parameters $\alpha$, $\theta$, and $\phi$. Then, we calculate the most general evolved state $\ket{\psi_\alpha}$ and its derivative $\partial_\alpha \ket{\psi_\alpha}$. We substitute into a formula for the QFI, then average over $\hat{\bm{n}}$.

We have expressed the rotation axis as 
$\hat{\bm{n}} = (\cos{\phi}\sin{\theta}, \sin{\phi}\sin{\theta}, \cos{\theta})$. In terms of this notation, the $U_\alpha$ generator
$A = - \hat{\bm{n}} \cdot \bm{\sigma}/2 $ has the form
\begin{equation} 
A  = \dfrac{1}{4}
\begin{pmatrix}
- \cos(\theta) & - e^{-i\phi} \sin(\theta) \\
- e^{i\phi} \sin(\theta) & \cos(\theta)
\end{pmatrix} \, .
\end{equation}
Hence the unitary $U_{\bm{\alpha}} = e^{i \alpha A}$ has the form
\begin{equation}
U_{\bm{\alpha}} =
  \begin{pmatrix}
\cos\left(\dfrac{\alpha}{2}\right) - i \cos(\theta) \sin\left(\dfrac{\alpha}{2}\right) 
& -  i e^{- i \phi}  \sin\left(\dfrac{\alpha}{2}\right) \sin(\theta)\\
- i e^{i\phi} \sin\left(\dfrac{\alpha}{2}\right) \sin(\theta) 
& \cos\left(\dfrac{\alpha}{2}\right) + i \cos(\theta) \sin\left(\dfrac{\alpha}{2}\right)
\end{pmatrix}
 \, .
\end{equation}

The most general input state has the form $\ket{\psi_0} = a \ket{z+} + b \ket{z-}$, with $a, b \in \mathbb{C}$ and $|a|^2 + |b|^2 = 1$. From this expression and the previous paragraph, we can calculate
$\ket{\psi_{\bm{\alpha}}}= U_{\bm{\alpha}} \ket{\psi_0}$
and its derivative, $\partial_{\alpha} \ket{\psi_{\bm{\alpha}}} $. 
We substitute into the QFI formula
$\mathcal{I}_{\alpha} = 4 \Re{\bra{\partial_{\alpha} \psi_{\bm{\alpha}}}\ket{\partial_{\alpha} \psi_{\bm{\alpha}}} - \left| \bra{\partial_{\alpha}  \psi_{\bm{\alpha}}} \ket{ \psi_{\bm{\alpha}}}\right|^2}$
[Eq.~1 in the main text]:
\begin{align}
  \mathcal{I}_{\alpha} 
  & = \dfrac{1}{2}\Big\{1 + 2|b|^2 - 2|b|^4 + (-1 + 6|b|^2 - 6|b|^4)\cos(2\theta) 
  \notag \\ & \hspace{1cm} 
  + 4b\Big[b(-1 + |b|^2)\cos(2\phi)\sin^2(\alpha) 
  + a(-1 + 2|b|^2)\cos(\phi)\sin(2\theta)\Big]\Big\} \, .
\end{align}
Next, we average $\phi$ and $\theta$ over their values on the unit sphere:
\begin{equation}
     \text{Avg}_{\hat{\bm{n}}}\big(\mathcal{I}_{\alpha} \big)  
     = \dfrac{1}{4\pi} 
     \int_{0}^{2 \pi}{\int_{0}^{\pi}
        \mathcal{I}_{ \bm{\alpha} } 
     \sin{\theta} \,  d \theta} \,  d \phi 
     = \dfrac{2}{3} \, .
\end{equation}

\subsubsection{The optimal ancilla-assisted entanglement-free input state saturates the lower bound in Eq.~\eqref{Eq_LowerBound}} \label{Sec_App_Entanglement_Free_Strategy}
The optimal family of ancilla-assisted entanglement-free sensors has the form
\begin{align} \label{Eq_General_Optimal_Classical_Ancilla_2}
    \rho_0 & = \dfrac{1}{3} \Big( \ketbra{z \pm}{z \pm} \otimes \ketbra{0}{0} + e^{i \phi_x}  \ketbra{x \pm}{x,\pm} \otimes \ketbra{1}{1} 
    +  e^{i \phi_y} \ketbra{y \pm}{y \pm} \otimes \ketbra{2}{2}\Big) \, .
\end{align}
As noted above, $\rho_0$ is a $1:1:1$ mixture of elements of three mutually maximally incompatible bases. The relative phases $e^{i \phi_x}$ and $e^{i \phi_y}$ are arbitrary.

Using Eq.~\eqref{Eq_General_Optimal_Classical_Ancilla_2}, we can write down different optimal input states: 
first, for each of $X$, $Y$ and $Z$, we can choose the eigenstate corresponding to the positive or negative eigenvalue.
For example, $\ketbra{z+}{z+}$ and $\ketbra{z-}{z-}$  work equally well. Second, the arbitrary phase factors $e^{i \phi_x}$ and $e^{i \phi_y}$ do not affect the QFIM. 
One such optimal input state is
\begin{align} \label{Eq_Optimal_Classical_Ancilla}
    \rho_0  & = \dfrac{1}{3} \left( \ketbra{z +}{z +} \otimes \ketbra{0}{0} + \ketbra{x +}{x +} \otimes \ketbra{1}{1} 
    +  \ketbra{y +}{y +} \otimes \ketbra{2}{2}\right) \, .
\end{align}
By Eq.~\eqref{Eq_Convexity_Simplified}, the corresponding QFIM is
\begin{equation}
  \mathcal{I}(\bm{\alpha} | \rho_{\bm{\alpha}} ) 
  = \dfrac{1}{3} \Big( \mathcal{I}_{{\ket{z +}}} 
  + \mathcal{I}_{{\ket{x +}}} 
  + \mathcal{I}_{{\ket{y +}}} \Big)  \, .
\end{equation}
$\mathcal{I}_{\ket{\hat{\bm{n}},+}}$ denotes the QFIM 
achievable with an input state equal to the eigenvalue-1 eigenstate of $\hat{\bm{n}}$. 
Calculating the expressions for $\mathcal{I}_{{\ket{z +}}}$, $\mathcal{I}_{{\ket{x +}}}$, and $\mathcal{I}_{{\ket{y +}}}$, we calculate the QFIM:
\begin{equation} \label{Eq_QFIM_classical_ancilla}
 \mathcal{I}(\bm{\alpha} | \rho_{\bm{\alpha}} ) = 
\begin{pmatrix}
\dfrac{2}{3} & 0 & 0 \\
0 & \dfrac{8}{3} \sin^2(\theta) \sin^2\left(\dfrac{\alpha}{2}\right) & 0 \\
0 & 0 & \dfrac{8}{3} \sin^2\left(\dfrac{\alpha}{2}\right)
\end{pmatrix} \, .
\end{equation}
This QFIM's inverse is
\begin{equation}
{\mathcal{I}(\bm{\alpha} | \rho_{\bm{\alpha}} )}^{-1} = 
    \begin{pmatrix}
    \dfrac{3}{2} & 0 & 0 \\
    0 & \dfrac{3}{8}\csc^2(\theta)\csc^2\left(\dfrac{\alpha}{2}\right) & 0 \\
    0 & 0 & \dfrac{3}{8}\csc^2\left(\dfrac{\alpha}{2}\right)
\end{pmatrix} \, .
\end{equation}
The $(1,1)$ entry equals $3/2$, the lower bound presented in Ineq.~\eqref{Eq_LowerBound}. Saturating this bound, the state~\eqref{Eq_Optimal_Classical_Ancilla} is an optimal entanglement-free probe.

\subsection{Entanglement-assisting sensing inspired by closed time-like curves} \label{Appendix_Bell_Sensing}

As before, we consider a probe undergoing an unknown rotation described by the operator $U_{\bm{\alpha}} = \exp \left(-i\alpha\hat{\bm{n}}\cdot\bm{\sigma} / 2 \right)$.
The rotation angle is $\alpha$, 
$\hat{\bm{n}} = \sin \theta \cos \phi \, \hat{\bm{x}} 
+ \sin \theta \sin \phi \, \hat{\bm{y}} 
+ \cos \theta \, \hat{\bm{z}}$ 
defines the rotational axis, and $\bm{\sigma} = (X, Y, Z)$ denotes a vector of Pauli operators. Relative to the $Z$ eigenbasis,
the eigenstates of $A = - \hat{\bm{n}}\cdot\bm{\sigma} /2$ are
\begin{align}
    \ket{a_+} = \dfrac{1}{\sqrt{1+ \tan^2{\dfrac{\theta}{2}}}} 
    \begin{pmatrix}
         - e^{- i \phi} \tan{\dfrac{\theta}{2}} \\ 1
    \end{pmatrix} 
    \quad \text{and} \quad
    \ket{a_-} = \dfrac{1}{\sqrt{1+ \cot^2{\dfrac{\theta}{2}}}}
    \begin{pmatrix}
         e^{- i \phi} \cot{\dfrac{\theta}{2}} \\ 1
    \end{pmatrix} \, .
\end{align}
Hence, $U_{\bm{\alpha}}$ can be expressed as 
\begin{align}
 e^{-i\alpha \mathbf{\sigma} \cdot\hat{\mathbf{n}} / 2} 
 & = e^{-i\alpha / 2} \ketbra{a_+}{a_+} + e^{-i\alpha / 2} \ketbra{a_-}{a_-} \\[0.25cm]
    & = \dfrac{ e^{- i\alpha / 2}}{1+ \tan^2{\dfrac{\theta}{2}}} \begin{pmatrix}
         - e^{- i \phi} \tan{\dfrac{\theta}{2}} \\ 1
    \end{pmatrix} \begin{pmatrix}
         - e^{i \phi} \tan{\dfrac{\theta}{2}} & 1
    \end{pmatrix} 
    + \dfrac{ e^{- i\alpha / 2}}{1+ \tan^2{\dfrac{\theta}{2}}}  \begin{pmatrix}
         e^{- i \phi} \cot{\dfrac{\theta}{2}} \\ 1
    \end{pmatrix}  \begin{pmatrix}
         e^{ i \phi} \cot{\dfrac{\theta}{2}} & 1
    \end{pmatrix} \\
        & = \dfrac{ e^{- i\alpha / 2}}{1+ \tan^2{\dfrac{\theta}{2}}} \begin{pmatrix}
          \tan^2{\dfrac{\theta}{2}} & - e^{- i \phi} \tan{\dfrac{\theta}{2}} \\ - e^{ i \phi} \tan{\dfrac{\theta}{2}} & 1
    \end{pmatrix} 
    + \dfrac{ e^{- i\alpha / 2}}{1+ \tan^2{\dfrac{\theta}{2}}}  \begin{pmatrix}
          \cot^2{\dfrac{\theta}{2}} &  e^{- i \phi} \cot{\dfrac{\theta}{2}} \\  e^{ i \phi} \cot{\dfrac{\theta}{2}} & 1
    \end{pmatrix} \\[0.25cm]
            & = e^{- i\alpha / 2}\begin{pmatrix}
          \sin^2{\dfrac{\theta}{2}} & - \dfrac{1}{2}e^{- i \phi} \sin{\theta} \\ - \dfrac{1}{2}e^{ i \phi} \sin{\theta} &\cos^2{\dfrac{\theta}{2}} 
    \end{pmatrix} 
    + e^{i\alpha / 2}\begin{pmatrix}
          \cos^2{\dfrac{\theta}{2}} &  \dfrac{1}{2}e^{- i \phi} \sin{\theta} \\  \dfrac{1}{2}e^{ i \phi} \sin{\theta} &\sin^2{\dfrac{\theta}{2}} 
    \end{pmatrix}  \\[0.25cm]
& = \begin{pmatrix}
          e^{- i\alpha / 2}  \sin^2{\dfrac{\theta}{2}} +   e^{ i\alpha / 2}  \cos^2{\dfrac{\theta}{2}}  & i \sin{\theta} \sin{\dfrac{\theta}{2}}   e^{ - i \phi} \\ i \sin{\theta} \sin{\dfrac{\theta}{2}}   e^{ i \phi} & e^{ i\alpha / 2}  \sin^2{\dfrac{\theta}{2}} +   e^{ - i\alpha / 2}  \cos^2{\dfrac{\theta}{2}}  
    \end{pmatrix} \\
    & \equiv \begin{pmatrix}
         a & b^* \\b & a^*
    \end{pmatrix}  \label{Eq_ab_def} \, . 
\end{align}
The final equation defines $a$ and $b$ implicitly.
The evolution of the probe and ancilla acts on their joint Hilbert space as
\begin{align}
     e^{-i\alpha \hat{\mathbf{n}} \cdot \mathbf{\sigma} / 2} \otimes \hat{\mathbb{1}} =    \begin{pmatrix}
          a  & 0 &  b^* & 0\\ 
          0 &  a & 0 &  b^* \\
          b & 0 &  a^* & 0  \\
          0 & b & 0 & a^*
    \end{pmatrix} \, .
\end{align}

Suppose that the probe and ancilla are prepared in an arbitrary Bell state $\ket{\Bell}$. Define the projector $\hat{\Pi}_\Bell \coloneqq \ketbra{\Bell}{\Bell}$. Recall that we aim to infer $\alpha$. An optimal measurement is entangling and has the form 
$\{ \hat{\Pi}_\Bell,  \hat{\mathbb{1}} - \hat{\Pi}_\Bell \}$.

For instance, suppose that the probe and ancilla begin in $\ket{\Phi^{+}}$. If 
$\hat{\Pi}_{\Phi^{+}} \coloneqq \ketbra{\Phi^{+}}{\Phi^{+}}$,
the optimal measurement is \{$\hat{\Pi}_{\Phi^{+}}$, 
$\hat{\mathbb{1}} - \hat{\Pi}_{\Phi^{+}}$\}. The possible outcomes obtain with the probabilities
\begin{align}
    P_0 & \coloneqq |\bra{\Phi^{+}}  U \ket{\Phi^{+}} |^2 
    = \Bigg| \dfrac{1}{2}
    \begin{pmatrix}
        1 & 0 & 0 & 1
    \end{pmatrix}
 \begin{pmatrix}
          a  & 0 &  b^* & 0\\ 
          0 &  a & 0 &  b^* \\
          b & 0 &  a^* & 0  \\
          0 & b & 0 & a^*
    \end{pmatrix}
        \begin{pmatrix}
        1 \\ 0 \\ 0 \\ 1
    \end{pmatrix} \Bigg|^2 
    = \Big| \dfrac{a + a^*}{2} \Big| ^2= \cos^2{\dfrac{\alpha}{2}} 
\end{align}
and $P_1 = 1- P_0 = \sin^2{\frac{\alpha}{2}} $. The resulting Fisher information is $\text{FI} = 1$ and equals the maximum QFI. 
Regardless of the choice of $\ket{\Bell}$, if the measurement is optimal, the FIM assumes the form
\begin{equation} 
\label{eq_I_help0}
I(\bm{\alpha} | \rho_{\bm{\alpha}} ) =
\begin{pmatrix}
1 & 0 & 0 \\
0 & 0 & 0 \\
0 & 0 & 0
\end{pmatrix} \, . 
\end{equation}
This FIM has the form in~\eqref{Eq_Special_QFIM}, with $\mathcal{I}_{\alpha} = 1$. 

The previous strategy provides no information about $\hat{\bm{n}}$:
$\mathcal{I}_{\hat{\bm{n}}} = 0$. Suppose that we wish to garner information about $\hat{\bm{n}}$, while optimally measuring $\alpha$. Our protocol's measurement basis can be modified to achieve this goal, we now show. First, we calculate the QFIM achievable with a general initial Bell state, $\ket{\Bell}$. This calculation bounds the attainable precision:
\begin{equation} \label{Eq_Optimal_Measurement_Basis}
\mathcal{I}(\bm{\alpha} | \rho_{\bm{\alpha}} ) =
\begin{pmatrix}
1 & 0 & 0 \\
0 & 4 \sin^2(\alpha/2) \sin^2{\theta} & 0 \\
0 & 0 & 4 \sin^2(\alpha /2) 
\end{pmatrix}  \, . 
\end{equation}
Now, we choose for the measurement basis to be \{$\ket{\Psi^{+}}$, 
$\ket{\Psi^{-}} $, $\ket{\Phi^{+}}$ , $\ket{\Phi^{-}}$\}. We calculate the FIM for this choice. The FIM, we show, equals the QFIM in Eq.~\eqref{Eq_Optimal_Measurement_Basis}. Therefore, this basis is the optimal basis for inferring about $\alpha$, while garnering information about $\hat{n}$.

Suppose that the input is the singlet state, $\ket{\Psi^{-}}$. One can perform similar calculations for input states equal to the other Bell states. 
The measurement's possible outcomes are obtained with probabilities 
\begin{align}
   P_{\Psi^+} & \coloneqq |\bra{\Psi^{+}}  U \ket{\Psi^{-}} |^2 = \cos^2(\theta)\sin^2\left(\frac{\alpha}{2}\right) \, , \\
   P_{\Psi^-} & \coloneqq |\bra{\Psi^{-}}  U \ket{\Psi^{-}} |^2  = \cos^2\left(\frac{\alpha}{2}\right) \, , \\
   P_{\Phi^+} & \coloneqq |\bra{\Phi^+}  U \ket{\Psi^{-}} |^2 = \cos^2(\phi)\sin^2(\theta)\sin^2\left(\frac{\alpha}{2}\right) \, , \; \: \text{and} \\
   P_{\Phi^-} & \coloneqq |\bra{\Phi^-}  U \ket{\Psi^{-}} |^2  = \sin^2(\phi)\sin^2(\theta)\sin^2\left(\frac{\alpha}{2}\right) \, .
\end{align}
We differentiate these probabilities with respect to $\alpha, \theta$, and $\phi$. We can then calculate the FIM using the formula
\begin{equation}
    I(\bm{\alpha})_{i,j} = \sum_{
        k \in \{ \Phi^{+}, \Phi^{-},
        \Psi^{+}, \Psi^{-} \} }
        {\dfrac{(\partial_i P_k) (\partial_j P_k) }{P_k ^2}} \, . 
\end{equation}
The result is
\begin{equation}
I(\bm{\alpha})  = \mathcal{I}(\bm{\alpha} | \rho_{\bm{\alpha}} )  = 
\begin{pmatrix}
1 & 0 & 0 \\
0 & 4 \sin^2(\alpha/2) \sin^2{\theta} & 0 \\
0 & 0 & 4 \sin^2(\alpha /2) 
\end{pmatrix}  ,
\end{equation}
which has the form of the optimal QFIM [Eq. \eqref{Eq_Optimal_Measurement_Basis}].

\end{widetext}


\begin{thebibliography}{10}
\bibitem{opticsnote}
\bibinfo{note}{Full Bell-basis measurements are not feasible in optics.
  However, our agnostic-sensing protocol requires only a measurement of whether
  the probe-and-ancilla system is in a singlet. This measurement can be
  achieved with Hong-Ou-Mandel
  interferometry~\cite{braunstein_bell_measurement_95}.}
\end{thebibliography}
\end{document}